\documentclass{article}

\usepackage{arxiv}
\usepackage[utf8]{inputenc} 
\usepackage[T1]{fontenc}    
\usepackage{hyperref}       
\usepackage{url}            
\usepackage{booktabs}       
\usepackage{amsfonts,amsmath,amsthm}       
\usepackage{nicefrac}       
\usepackage{microtype}      
\usepackage{lipsum}		
\usepackage{graphicx}
\usepackage{natbib}
\usepackage{doi}
\usepackage[skins]{tcolorbox}

\newtheorem{definition}{Definition}
\newtheorem{theorem}{Theorem}
\newtheorem{lemma}{Lemma}
\newtheorem{proposition}{Proposition}
\newtheorem{corollary}{Corollary}
\newtheorem{observation}{Observation}

\title{On the Arrow of Inference}

\author{ \href{https://orcid.org/0000-0003-2067-2763}{\includegraphics[scale=0.06]{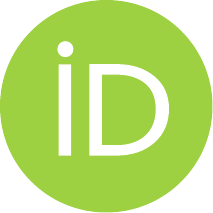}\hspace{1mm}Xin Li}\thanks{This work was partially supported by NSF IIS-2401748 and BCS-2401398. The author has used ChatGPT models to assist the development of theoretical ideas and the proof of theorems presented in this paper.} \\
	Department of Computer Science\\
	University at Albany\\
	Albany, NY 12222 \\
	\texttt{xli48@albany.edu} 
}

\renewcommand{\shorttitle}{\textit{arXiv} Template}

\hypersetup{
pdftitle={A template for the arxiv style},
pdfsubject={q-bio.NC, q-bio.QM},
pdfauthor={David S.~Hippocampus, Elias D.~Striatum},
pdfkeywords={First keyword, Second keyword, More},
}

\begin{document}
\maketitle

\begin{abstract}
Just as the arrow of time underpins the foundations of physics, the arrow of inference emerges as a fundamental organizing principle of cognition, governing the directional flow of information in perception, action, and memory. We introduce the Context-Content Uncertainty Principle (CCUP) as a unifying theoretical framework that formalizes this asymmetry between high-entropy context and low-entropy content, and show how inference proceeds by forming cycles that align context with content through selective, bidirectional interaction.

Within this framework, cycle formation resolves the Information Bottleneck (IB) in Optimal Transport (OT) by coordinating bottom-up contextual disambiguation with top-down content reconstruction. This dual mechanism, implemented as a Rao-Blackwellized inference strategy, is reflected neurobiologically in the dorsal (context) and ventral (content) streams, which operate in cyclical coordination to minimize joint uncertainty across hierarchical representations. Crucially, this local cycle completion is extended into chain formation, wherein each cycle serves as a link in a temporally structured inference trajectory, or \emph{memory chain}, that enables the system to simulate goals, evaluate counterfactuals, and refine internal models over time. Spatial bootstrapping through ventral-dorsal separation and temporal bootstrapping through the perception–action cycle jointly support progressive refinement of sensorimotor models through environmental interaction. Moreover, the sleep–wake cycle extends CCUP across timescales, transforming episodic memory into semantic knowledge through recursive consolidation along both temporal chains and spatial hierarchies.

We further develop a representational extension of CCUP, wherein each level performs delta-seeded inference: internally generated content seeds unfold outward through structured diffusion, constrained by top-down priors and external context. This mechanism circumvents the curse of dimensionality by confining inference to low-dimensional manifolds aligned with goal-relevant features. Building on this structure, we propose that language emerges as a social extension of the inference cycle: a symbolic system for transporting latent content across minds, resolving intersubjective IB in OT through shared anchoring. 
Together, these results position CCUP as a foundational principle underlying the arrow of inference in both individual cognition and collective intelligence.
\end{abstract}


\section{Introduction: The Arrow of Inference}
\label{sec:1}

In physics, the \emph{arrow of time} describes the unidirectional evolution toward greater entropy, grounding our understanding of causality and irreversibility \cite{zeh2007physical}. By analogy, we propose an \emph{arrow of inference} in cognition: a directional flow from ambiguous, high-entropy context to specific, low-entropy content, enabling intelligent systems to reduce uncertainty about latent causes \cite{pearl2009causality}. Unlike the thermodynamic arrow \cite{friston2010free,gershman2019what}, the arrow of inference arises from an asymmetry in information flow: contextual cues are stochastic and ambiguous, while content representations are deterministic and behaviorally relevant \cite{bar2004visual,clark2013whatever}. This directionality is shaped by cognitive architecture and task demands, reflecting a structural bias from uncertainty toward specificity.

To formalize the asymmetry between context and content, we introduce the \textbf{Context–Content Uncertainty Principle} (CCUP), which posits a trade-off between the entropy of contextual variables and the specificity of content representations \cite{li2025CCUP}. CCUP frames inference as a cyclical process: bottom-up context disambiguation guides content binding, while top-down content shapes context reconstruction/generation. This bidirectional alignment resolves the information bottleneck (IB) \cite{tishby2000information} in optimal transport (OT) \cite{villani2009optimal} by replacing compression with bottom-up conditional inference and relevance with top-down manifold-constrained reconstruction. These processes resemble a Rao-Blackwellization strategy \cite{rao1992information,blackwell1947conditional}, in which context conditions content to reduce uncertainty, while content enables generalization across contexts. Rather than separating encoding and decoding, CCUP supports co-evolving inference cycles into \emph{memory chains} that dynamically align latent representations \cite{dayan1995helmholtz,buzsaki2006rhythms}. In this view, the bottleneck becomes a scaffold: uncertainty is resolved not by reduction, but by dynamic alignment and iterative reconstruction \cite{hinton1995wake}.

The CCUP framework grounds the arrow of inference in brain architecture through \emph{spatial bootstrapping} between the ventral and dorsal streams: the dorsal stream disambiguates perception via top-down contextual disentangling, while the ventral stream refines bottom-up content binding \cite{ungerleider1994and,rao1999predictive}. This interaction implements a spatial form of Rao-Blackwellization, iteratively aligning context and content to reduce joint uncertainty \cite{friston2008hierarchical}. Extending this principle in time, the \emph{perception–action cycle} functions as \emph{temporal bootstrapping}, where action reshapes the environment to improve perceptual inference through recursive engagement \cite{fuster2004upper}. In this loop, bottom-up perception gathers uncertain evidence, while top-down action reshapes the generative context, forming a closed cycle that iteratively reduces epistemic uncertainty \cite{hullermeier2021aleatoric}. Beyond bidirectional inference, CCUP also accounts for \emph{memory consolidation} across the sleep–wake cycle: episodic experiences \cite{tulving2002episodic} are replayed and reorganized into semantic structures, transforming ambiguity into abstract knowledge over extended timescales \cite{diekelmann2010memory,stickgold2005sleep}. 

Building on this foundation, we propose a \emph{representational extension of CCUP}, where each cognitive level performs \emph{delta-seeded diffusion} \cite{bansal2023cold}: inference begins from a low-entropy latent seed (e.g., an internally generated goal) and unfolds selectively along structured manifolds guided by top-down priors and bottom-up evidence \cite{rombach2022high}. This strategy circumvents the curse of dimensionality by constraining inference to goal-relevant subspaces, transforming the bottleneck into a scaffold for abstraction \cite{bengio2013representation}. Unlike Gibbs distribution \cite{geman1984stochastic}, where inference requires expensive global normalization via the partition function, delta-seeded diffusion circumvents this by initializing inference with sharply peaked priors, favoring efficient local updates. In addition to biological plausibility (e.g., supported by the discovery of place cells \cite{moser2008place} and concept cells \cite{quiroga2005invariant}), delta-seeded diffusion significantly reduces computational complexity by avoiding high-dimensional integrals.
The extended CCUP mechanism underlies the emergence of symbolic cognition: \emph{language} arises as a social extension of delta-seeded inference, transmitting latent content across minds via shared symbols that initiate inference cycles in others \cite{pickering2004toward}. 
In sum, the arrow of inference captures a deep regularity underlying cognition and intelligence: {\bf selective, dynamic, and asymmetrical alignment between content and context across scales, modalities, and agents}. 

\section{Theoretical Foundation: Context–Content Uncertainty Principle (CCUP)}
\label{sec:2}

Cognition, at its core, is a process of resolving uncertainty \cite{robertson1929uncertainty}. Every act of perception, action, or memory retrieval involves mapping ambiguous sensory or internal signals onto specific, behaviorally relevant interpretations. We propose CCUP as a foundational framework for understanding this process. CCUP asserts that context and content are informationally asymmetric \cite{anderson1972more}: although context variables may occupy a lower-dimensional sensory space, they exhibit high entropy due to ambiguity and observational limits; in contrast, content variables often span a broader semantic space but maintain low entropy due to goal-specific selection \cite{friston2005theory}. Thus, CCUP characterizes inference not as compression, but as a {\em directional process} akin to inverse diffusion \cite{song2021score}: from entropic context to selective, structured content. This asymmetry drives the directional flow of inference - {\em one can use context to predict content reliably, but not content to infer context unambiguously}. 

\subsection{Asymmetry Creates Directional Information Loss: a CCUP-Based Analysis}

Let \( \Psi \) represent high-entropy contextual variables (e.g., sensory observations or task goals) and \( \Phi \) represent low-entropy content variables (e.g., object identity or action choice). The task of inference is to estimate \( \Phi \) given \( \Psi \), but due to the ambiguity and variability of \( \Psi \), a direct mapping often leads to loss of specificity or poor generalization. For any cognitive task, an agent must infer stable latent causes from aliased sensory observations \cite{clark2013whatever}. This inferential directionality, from uncertainty to specificity, is not arbitrary but structured by cognitive architecture and task demands. More fundamentally, it reflects an inherent asymmetry in the flow of information: {\em contextual variables tend to be entropic and ambiguous}, while {\em content variables are specific and actionable} \cite{bar2004visual}. The following lemma summarizes the above observation.

\begin{lemma}[Asymmetry of Inference in Information-Rich Systems]
Let \( \Psi \) denote a high-entropy contextual variable, and let \( \Phi \) denote a low-entropy, stable content variable inferred from \( \Psi \). The CCUP formalizes an inherent asymmetry in inference between these variables, reflected in their conditional entropies.
In the forward direction, the mapping \( \Psi \rightarrow \Phi \) involves disambiguating aliased contextual input to infer a precise latent cause. This process is underdetermined, yielding nonzero conditional entropy \cite{cover1999elements}:
$H(\Phi \mid \Psi) > 0$.
In the reverse direction, the mapping \( \Phi \rightarrow \Psi \) requires generating/simulating a range of compatible contextual instances from a compact content representation. This generalization introduces additional uncertainty: $H(\Psi \mid \Phi) \gg H(\Phi \mid \Psi)>0$.
\end{lemma}

This asymmetry of inference causes the flow \( \Psi \rightarrow \Phi \rightarrow \Psi \) to be \emph{non-conservative} in terms of mutual information \cite{shannon1948mathematical}:
$I(\Psi; \Phi)=H(\Psi)- H(\Psi \mid \Phi)< H(\Psi), \quad \text{and} \quad H(\Psi \mid \Phi) > 0$.
This directional loss is not due to suboptimal learning but is \emph{structurally unavoidable} given the entropy gradient between context and content. Asymmetry in inference is not merely local (directional), but is bounded globally by the mismatch in marginal entropies. Since both directions \( \Psi \rightarrow \Phi \) and \( \Phi \rightarrow \Psi \) carry irreducible uncertainty, the sum of these uncertainties must respect the minimum divergence in information content. Formally, we have (its proof can be found in Appendix A).

\begin{theorem}[Context-Content Uncertainty Principle (CCUP)]
If we define uncertainty arising from contextual disambiguation 
and contextual reconstruction by $H(\Phi \mid \Psi)$ and
$H(\Psi \mid \Phi)$, respectively, then we have 
\begin{equation}
    H(\Phi \mid \Psi) + H(\Psi \mid \Phi) \geq \left| H(\Phi) - H(\Psi) \right|.
    \label{eq:1}
\end{equation}
\end{theorem}

\noindent
\textbf{Remark.}
This inequality formalizes a trade-off: increasing specificity by reducing \( H(\Phi \mid \Psi) \) inherently increases the difficulty of reconstructing context from the latent representation by increasing \( H(\Psi \mid \Phi) \), and vice versa. CCUP therefore imposes a fundamental asymmetry in how context and content contribute to representation learning and inference \cite{bengio2013representation}: precise context may still yield ambiguous content, while stable content representations must generalize over ambiguous context. 

\subsection{Dynamic Alignment of Content and Context Representation}

The inherent asymmetry in entropy between context and content variables makes information loss during optimal transport (OT) \cite{villani2009optimal} unavoidable, creating an information bottleneck (IB) \cite{tishby2000information}. Theorem 1 implies that \emph{one-way transport} cannot preserve mutual information: encoding context into content inherently discards information unless explicitly corrected via reconstructive feedback. The only way to overcome this bottleneck is through dynamic alignment: an inference cycle that co-evolves content and context representations via top-down and bottom-up passes. This insight informs both theoretical architectures (e.g., variational autoencoders \cite{kingma2014auto}, Helmholtz machines \cite{dayan1995helmholtz}) and biological systems (e.g., ventral-dorsal loops \cite{ungerleider2000mechanisms}, sleep-based memory consolidation \cite{stickgold2005sleep}), where \emph{recurrent inference cycles are necessary} to counteract irreversible information loss.

\paragraph{How IB in OT Arises from Directional Asymmetry under CCUP.}
Under the framework of OT \cite{villani2009optimal}, a high-entropic contextual distribution \( \mu(\Psi) \) must be mapped onto a lower-entropic content distribution \( \nu(\Phi) \), often via a latent variable \( Z \). Based on CCUP, the asymmetry between \( H(\Psi) \) and \( H(\Phi) \) induces directional information loss. Therefore, \emph{asymmetry creates an arrow}: the inference process must compensate for this directional loss by cycling- i.e., refining \( \Phi \) through repeated bottom-up passes and updating \( \Psi \) through top-down reconstruction. Only through such \emph{cycle formation} can the system mitigate information loss and stabilize representations.
Any latent encoding \( \Psi \to Z \) that attempts to encode context for relevance to \( \Phi \) must confront an IB: \( Z \) cannot simultaneously preserve enough information for both content specificity and context recovery. Thus, the classical IB in OT is not merely a constraint on mutual information but a structural consequence of asymmetry in uncertainty. Instead of one-shot encoding, CCUP reframes this challenge as a continual coordination problem between two representations that must be dynamically aligned \cite{friston2008hierarchical}. We summarize the strategy of dynamic alignment into the following lemma.

\begin{lemma}[Entropy Asymmetry Implies Bottleneck in Optimal Transport]
Let $\Psi$ and $\Phi$ be random variables representing high-entropy context and low-entropy content, respectively. Suppose the joint distribution $p(\Psi, \Phi)$ is transported via a coupling $\gamma \in \Pi(p(\Psi), p(\Phi))$. Under CCUP, we observe:
$H(\Psi) \gg H(\Phi)  \Rightarrow I(\Psi; \Phi) < H(\Psi)$,
which implies that any transport from $p(\Psi)$ to $p(\Phi)$ incurs an \emph{information bottleneck}.
Moreover, static transport mappings $T: \Psi \mapsto \Phi$ cannot preserve mutual information. To recover representational fidelity, the system must employ \textbf{dynamic alignment}: a cyclic process alternating between bottom-up inference $p(\Phi|\Psi)$ and top-down reconstruction $p(\Psi|\Phi)$.
\end{lemma}


\paragraph{CCUP-based Design Principle for Cycle Formation.}
In classical formulations of IB, OT faces the challenge of mapping a high-entropy contextual distribution \( \mu(\Psi) \) to a lower-entropy content distribution \( \nu(\Phi) \) via a latent variable \( Z \). Traditional wisdom frames this as a one-shot encoding problem: how to encode \( \Psi \) into \( Z \) while preserving relevance to \( \Phi \), which introduces an IB due to the entropy gap \( H(\Psi) \gg H(\Phi) \).
CCUP offers a fundamentally different perspective on introducing a feedback loop to form a dynamic cycle. 
First, it reframes the problem as one of \emph{coordination} between two asymmetric representations that must be dynamically aligned. Rather than discarding information, inference under CCUP involves disambiguating \( \Psi \) through bottom-up encoding/conditioning \( p(\Phi \mid \Psi) \), and reconstructing \( \Psi \) from \( \Phi \) via top-down decoding/reconstruction \( p(\Psi \mid \Phi) \). These are not opposing or dialectical constraints but mutually supportive operations that iteratively align bilateral (content and context) representations through bidirectional inference \cite{parr2017uncertainty}. 

Second and more crucially, CCUP resolves the bottleneck through \emph{cycle formation}, which transforms the directional arrow of inference into a convergent loop that distributes inference across time. 
The cyclic system turns the entropy asymmetry between context and content into a temporal resource, bootstrapping stability through dynamic consistency \cite{friston2017graphical}.
More specifically, CCUP motivates the following cyclic inferential architecture: a \emph{bottom-up pass} disambiguates context through Rao-Blackwellization-based \emph{conditioning}   \cite{rao1992information,blackwell1947conditional} or specification \cite{zhang2017understanding} where \( p(\Phi \mid \Psi) \) is sharpened by exploiting structure in \( \Psi \); this is followed by a \emph{top-down pass} that reconstructs or predicts \( \Psi \) from inferred content \( \Phi \), modeling \( p(\Psi \mid \Phi) \). Note that top-down reconstruction often involves \emph{aliasing} or generalization \cite{bousquet2002stability} - that is, reusing stable content representations across multiple contextual situations, enabling abstraction and generalization. In sum, inference becomes a bidirectional loop that minimizes the joint uncertainty \( H(\Psi, \Phi) \) while maximizing mutual information \( I(\Psi; \Phi) \), aligning representations through repeated cycles \cite{tishby2015deep}.

\begin{proposition}[Cycle Formation under CCUP]
Let \( \Phi \) denote a low-entropy latent content representation, and \( \Psi \) denote a high-entropy contextual observation. Under CCUP, nature completes the cycle
$\Psi \xrightarrow{\text{encoding}} \Phi \xrightarrow{\text{decoding}} \hat{\Psi} \xrightarrow{\text{update}} \Phi'$
such that:
1) \textbf{Content encoding:} \( \Phi \) is an encoded representation of \( \Psi \) satisfying \( H(\Phi) \ll H(\Psi) \); 2) \textbf{Contextual decoding:} A generative process \( G \) expands \( \Phi \) into a prediction \( \hat{\Psi} = G(\Phi) \), leading to prediction error: A comparator computes the discrepancy \( \delta = d(\Psi, \hat{\Psi}) \); and
    3) \textbf{Content update:} The representation is updated as \( \Phi' = \Phi - \eta \nabla_\Phi d(\Psi, \hat{\Psi}) \), reducing uncertainty via gradient descent or biological analogy (e.g., synaptic plasticity).
If this cycle is iterated with bounded reconstruction error \( d(\Psi, \hat{\Psi}) \leq \epsilon \), the process converges to a fixed point where content and context are dynamically aligned, and mutual information \( I(\Phi; \Psi) \) is maximized.
\end{proposition}

\noindent
\textbf{Remark.}
The above proposition extends and reinterprets Friston’s predictive coding \cite{friston2005theory} by emphasizing the inherent asymmetry between low-entropy content and high-entropy context. While predictive coding theory \cite{friston2006free} models the brain as minimizing prediction error via top-down priors and bottom-up residuals, CCUP frames inference as a dynamic cycle that aligns context and content through alternating disambiguation and reconstruction. Crucially, CCUP formalizes inference as a process of entropy transformation, selectively unfolding high-entropic context from specific content rather than correcting prediction errors, as we will elaborate next. 

\subsection{Bidirectional Interaction Breaks the Arrow of Inference}

Classical formulations of the IB in OT aim to map a high-entropy contextual distribution \( \mu(\Psi) \) to a lower-entropy content distribution \( \nu(\Phi) \) via a latent variable \( Z \), while minimizing transport cost and preserving relevance. This is traditionally viewed as a compression-relevance trade-off, with the encoding \( p(Z \mid \Psi) \) constrained to retain as much information about \( \Phi \) as possible.
However, CCUP reveals that this bottleneck emerges not from compression per se, but from an underlying \emph{asymmetry of directional information loss}. 
To resolve this asymmetry, CCUP proposes that inference must proceed through a \emph{cycle} composed of:
\begin{itemize}
    \item \textbf{Bottom-up encoding} \( p(\Phi \mid \Psi) \) that performs contextual disambiguation, minimizing \( H(\Phi \mid \Psi) \),
    \item \textbf{Top-down decoding} \( p(\Psi \mid \Phi) \) that performs contextual reconstruction, minimizing \( H(\Psi \mid \Phi) \).
\end{itemize}

These processes must be coordinated through \emph{iterative bidirectional inference} \cite{dayan1995helmholtz,hinton1995wake}, such that the forward and backward mappings form a cycle that aligns representations across inference passes. Formally, we have 

\begin{theorem}[Cycle Consistency Eliminates the IB in OT]
Let \( \Psi \) and \( \Phi \) be random variables representing context and content, respectively, with associated conditional distributions \( p(\Phi \mid \Psi) \) and \( p(\Psi \mid \Phi) \). Suppose that inference proceeds via an iterative cycle:
$\Psi \rightarrow \Phi \rightarrow \Psi$,
where each pass updates representations to reduce uncertainty. If this bidirectional inference process converges to a fixed point such that the forward and backward transport maps become cycle-consistent, i.e., $\Psi \approx \mathbb{E}_{p(\Psi \mid \Phi)}[\Psi], \quad \Phi \approx \mathbb{E}_{p(\Phi \mid \Psi)}[\Phi],$
then the conditional entropies vanish: $H(\Phi \mid \Psi) \to 0, \quad H(\Psi \mid \Phi) \to 0,$
and mutual information is fully recovered: $I(\Psi; \Phi) = H(\Psi) = H(\Phi).$
Thus, the IB is eliminated and OT is achieved through dynamic alignment rather than one-time encoding.
\end{theorem}

\noindent
\textbf{Remark.}
The proof of the above theorem is referred to in Appendix B. This theorem demonstrates that IB in OT is not an inherent limitation of representation capacity but a consequence of misalignment between asymmetric variables. Under CCUP, this misalignment can be overcome by cycle formation that iteratively aligns representations through bottom-up disambiguation and top-down reconstruction \cite{rao1999predictive}. Once the forward and backward inference paths (encoding and decoding) become consistent, the system achieves mutual certainty and fully recovers the transport map between \( \Psi \) and \( \Phi \) without information loss. Next, we elaborate on the design principles underlying cyclic inference.

\paragraph{Context Disambiguation via Selective Inference (Rao-Blackwellization).}
CCUP reframes the task of inference as \emph{dynamic alignment} between asymmetrical representations. Specifically, bottom-up encoding is understood as a process of \emph{contextual disambiguation} through selective inference, wherein content representations \( \Phi \) are refined by conditioning on rich, high-dimensional contextual inputs \( \Psi \). This process is formally analogous to Rao-Blackwellization \cite{rao1992information,blackwell1947conditional}, in which variance in the inference of \( \Phi \) is reduced by exploiting sufficient statistics in \( \Psi \) (i.e., {\em conditioning reduces variance}). Rather than compressing away the complexity of context, CCUP emphasizes leveraging it to sharpen and stabilize content through selective inference or conditioning. This approach preserves relevant information while reducing epistemic uncertainty by anchoring inference in the structure provided by context \cite{friston2008hierarchical, parr2019generalised}. A similar idea, named ``pre-wired bias'' or ``inductive bias'' \cite{goyal2022inductive}, was discussed for resolving the bias-variance dilemma \cite{geman1992neural}. 

\paragraph{Context Reconstruction Through Aliasing for Generalization.}
In the reverse direction, CCUP replaces the classical notion of \emph{relevance}, preserving only predictive bits about the output, with a more generative notion of \emph{context reconstruction}. Once stable content representations are inferred, they are reused or \emph{aliased} across diverse contextual instances during top-down decoding. This aliasing mechanism is not a flaw but a powerful enabler of generalization: it allows the system to map multiple high-entropy contextual inputs onto a common latent representation, thereby supporting abstraction across variation \cite{bar2004visual}. From the perspective of conditional entropy, this enables the model to tolerate \( H(\Psi \mid \Phi) > 0 \) while still recovering functional reconstructions of context. In this way, CCUP formalizes generalization as the reuse of compact content structures across ambiguous and overlapping contexts, a view that aligns with empirical observations of representational invariance in both biological and artificial systems \cite{zhang2017understanding}.

\section{Chain Representation of Context and Content for Bidirectional Inference}
\label{sec:3}

A fundamental new insight underlying our CCUP-based approach is that they both boil down to the positional encoding problem in the latent space from the perspective of {\em goal-directed behavior} \cite{ajzen1986prediction}. Both goal and memory are computational tricks invented by nature for efficiently solving Minsky's search problem \cite{minsky1961steps}. The key idea is to combine Bellman's principle of optimality \cite{bellman1966dynamic} (i.e., global-to-local via dynamic programming) with Friston's principle of causal/active inference \cite{friston2017active} (i.e., local-to-global causation via goal simulation). Formally, we have the following proposition.

\begin{proposition}[Dynamic Programming via Temporal Bootstrapping]
    Goal simulation for goal-directed behavior can be framed as a form of bootstrapped dynamic programming, where the system incrementally constructs value estimates and policies not from fixed external rewards, but through internally generated goals and predictive models within a spatiotemporal context specified by the memory chain.
\end{proposition}

\textbf{Remark.} 
While cycle formation under CCUP enables the alignment of content and context through iterative inference \cite{friston2017graphical}, it primarily addresses localized uncertainty minimization within closed loops. However, real-world cognition often unfolds as a sequence of evolving states where inference must be extended over temporally and causally structured trajectories \cite{schacter2007cognitive}. To capture this unfolding structure, \emph{cycle completion must be generalized into chain formation}, allowing the system to simulate goals and actions over extended horizons. 

\subsection{Temporal Bootstrapping: From Cycle Completion to Chain Formation}

In this formulation, each link in the chain corresponds to a partially resolved cycle, locally minimizing uncertainty while contributing to the global construction of a coherent trajectory. This extension supports \emph{bootstrapped dynamic programming} \cite{sutton1998reinforcement}, where internally generated goals propagate value estimates forward in time, and policy refinement emerges through recursive inference grounded in a \emph{memory chain} \cite{wang2018prefrontal}. Thus, chain formation provides the organizational substrate for spatiotemporal goal simulation, enabling the agent to dynamically construct and navigate latent plans in the absence of fixed external rewards \cite{botvinick2020deep}. We start with the following construction.

\begin{definition}[Memory Chain]
Let $\{Z_t\}_{t=1}^T$ be a sequence of latent variables representing internal memory states indexed over discrete time steps $t = 1, \dots, T$. A \emph{memory chain} is a structured trajectory through latent space such that:

\begin{enumerate}
    \item Each $Z_t$ is conditionally dependent on both the preceding state $Z_{t-1}$ and an associated context variable $X_t$, forming the joint model:
    $p(Z_{1:T}, X_{1:T}) = p(Z_1) \prod_{t=2}^T p(Z_t \mid Z_{t-1}, X_t)$,

    \item The inference over the chain involves iterative bidirectional updates:    
   $q(Z_t) \propto \psi_t(Z_t) \prod_{s \in \mathcal{N}(t)} \phi_{s \to t}(Z_t),$    
    where $\psi_t$ denotes the local potential incorporating contextual evidence, and $\phi_{s \to t}$ are global messages passed from neighboring latent states $Z_s$.
    
    \item The memory chain supports approximate inference by minimizing a variational free energy objective over time:
    $\mathcal{F}[q] = \sum_{t=1}^T \mathbb{E}_{q(Z_t)}[-\log p(X_t \mid Z_t)] + \text{KL}(q(Z_{1:T}) \Vert p(Z_{1:T})).$
    
\end{enumerate}

\end{definition}  

\textbf{Remark.} A memory chain represents a temporally extended, entropy-constrained trajectory of internal states, where each link in the chain performs partial cycle completion between content and context. It generalizes localized cycle-based inference into spatiotemporal alignment across multiple levels of abstraction. Several empirical observations support the biological plausibility of memory chain. For example, the phase of sloped recession refers to the systematic advancement (precession) of spike timing relative to theta phase as an animal traverses a spatial field (pp. 320, \cite{buzsaki2006rhythms}). The slope of this phase shift encodes spatial or episodic structure and reflects intrinsic oscillatory dynamics and network interactions. The temporal receptive window of neural circuits increases hierarchically along the sensory-to-association cortex axis, enabling cumulative integration of past information over increasing timescales \cite{hasson2015hierarchical}. Heuristically, local potentials (context $\psi$) constrain immediate inference; while global potentials (content $\phi$) represent stable representations or structured priors embedded into the memory chain. A crucial missing link is the bridge between local and global potentials. 

\paragraph{Cloning as a Bridge Between Cycles.} 
The CCUP framework posits that inference operates through cycles that iteratively align high-entropy context variables \( \Psi \) with low-entropy content variables \( \Phi \), minimizing joint uncertainty through bidirectional inference. While individual cycles resolve local ambiguities between context and content, intelligent cognition requires coherence across temporally or spatially extended experience.
We propose that such global coherence arises through \textbf{chain formation}, defined as the structured \emph{concatenation of inference cycles}:
\[
(\Psi_1 \leftrightarrow \Phi_1) \rightarrow (\Psi_2 \leftrightarrow \Phi_2) \rightarrow \cdots \rightarrow (\Psi_n \leftrightarrow \Phi_n)
\]
Each link in the chain corresponds to a local inference cycle, where \( \Phi_k \) is inferred from \( \Psi_k \) and reused or refined in \( \Psi_{k+1} \), possibly via a {\em cloning mechanism} \cite{george2021clone} that allows semantic continuity across varying contexts.
To ensure continuity and generalization, a cloned representation of content \( \Phi_k \) is propagated forward, seeding the next cycle by anchoring it to previously stabilized latent variables. This enables the system to bootstrap learning across steps, supporting goal persistence, memory chaining, and temporal abstraction.
In the spatial domain, chaining cycles corresponds to navigating across locations where each landmark (i.e., delta seeding \cite{song2021score, blundell2015weight}) anchors the next perceptual update. In the temporal domain, it supports memory consolidation, where episodic events are linked via shared content variables, forming a coherent narrative structure. In either case, chain formation implements a form of \emph{latent navigation} \cite{ho2022classifierfree,locatello2019challenging}, where the agent traverses an abstract representational space through sequentially coupled inference cycles.

\paragraph{From Inference Cycles to Cognitive Trajectories.}  
Chain formation transforms CCUP’s local cycles into global trajectories: structured paths through latent space that preserve semantic consistency while accommodating contextual change. It also explains how cognitive systems sustain identity, intention, and learning across time: not through persistent activation, but through structured reactivation of cycle-aligned content across a chain of inference steps.

\begin{definition}[Inference Chains as Structured Latent Trajectories]
Let each local cycle \( (\Psi_k \leftrightarrow \Phi_k) \) minimize joint uncertainty via CCUP. Then a chain of such cycles, with cloned propagation of \( \Phi_k \) to seed \( \Phi_{k+1} \), forms a structured trajectory in latent space that enables generalization, abstraction, and temporally extended cognition.
\end{definition}

With the above definition, we can connect chain formation as structured trajectories in the latent space with graph-based inference such as loopy belief propagation (LBP) \cite{weiss2001optimality}.

\begin{proposition}[Chain Formation as Structured Loopy Belief Propagation]
Let each cycle in a CCUP-aligned system be defined by a bidirectional inference process between a high-entropy contextual variable \( \Psi_k \) and a low-entropy content variable \( \Phi_k \), minimizing conditional entropies \( H(\Phi_k \mid \Psi_k) \) and \( H(\Psi_k \mid \Phi_k) \). Define a \emph{chain} as a sequence of such cycles indexed by \( k = 1, \dots, n \), where each content variable \( \Phi_k \) is cloned into the next cycle as a latent prior via a structural constraint \( \psi_k(\Phi_k, \Phi_{k+1}) \).
Then, the joint inference process over the chain:
$p(\Phi_{1:n}, \Psi_{1:n}) \propto \prod_{k=1}^{n} \phi_k(\Phi_k, \Psi_k) \cdot \prod_{k=1}^{n-1} \psi_k(\Phi_k, \Phi_{k+1})$
approximates LBP on a factor graph with local cycle potentials \( \phi_k \) and inter-cycle transition potentials \( \psi_k \). Each cycle performs message passing via local inference, and the chaining of these cycles implements a global approximate inference procedure that refines marginals through repeated alignment.
\end{proposition}

\paragraph{Chain Formation in Resolving the IB for OT.} 
It follows from Proposition 3 that chain formation under CCUP corresponds to structured loopy BP, where inference converges through recurrent, context-sensitive updates across a loopy graphical structure.
While cycle formation resolves directional uncertainty by aligning representations across inference passes, it assumes that the latent space has sufficient capacity to capture all relevant contextual distinctions \cite{kingma2014auto}. However, multiple high-entropy contexts $\Psi_i$ may map to the same or overlapping content representations $\Phi$, leading to aliasing and degraded reconstructions. This is where \emph{cloning}, the duplication or branching of content representations to account for multiple context-dependent variants of the same latent concept, becomes essential \cite{george2021clone}. 

\begin{proposition}[Chain Formation and Cloning Resolve the IB in OT]
Let $\Psi$ denote a high-entropy contextual variable and $\Phi$ a low-entropy content variable. Let an inference system seek to transport $\mu(\Psi)$ to $\nu(\Phi)$ via a latent map while preserving mutual information and minimizing directional uncertainty.
Under the framework of CCUP, the following two mechanisms jointly eliminate the IB in OT:

\begin{itemize}
  \item \textbf{Chain Formation:}  
  If the system constructs a temporally ordered inference trajectory  
$\Psi_1 \rightarrow \Phi_1 \rightarrow \Psi_2 \rightarrow \Phi_2 \rightarrow \cdots \rightarrow \Psi_T \rightarrow \Phi_T$,  
such that each pair $(\Psi_t, \Phi_t)$ undergoes local cycle completion and the sequence converges toward \emph{global chain consistency}, then conditional uncertainties across all steps are minimized:
$H(\Phi_t \mid \Psi_t) \rightarrow 0 \quad \text{and} \quad H(\Psi_{t+1} \mid \Phi_t) \rightarrow 0 \quad \forall t$,  
yielding recovery of \emph{maximal mutual information across the chain}:
$I(\Psi_{1:T}; \Phi_{1:T}) = H(\Psi_{1:T}) = H(\Phi_{1:T})$.

  \item \textbf{Cloning Trick:}  
  If a single content variable $\Phi$ cannot adequately represent multiple distinct contextual instances $\Psi_1, \Psi_2, \ldots, \Psi_n$, then the system must introduce cloned content variables $\Phi^{(1)}, \Phi^{(2)}, \ldots, \Phi^{(n)}$ such that  
  $\forall i,\ \Phi^{(i)} \approx \Phi$ but $H(\Psi_i \mid \Phi^{(i)}) \ll H(\Psi_i \mid \Phi)$,  
  thereby preserving contextual distinctions without collapsing content representations.
\end{itemize}

Together, these two mechanisms convert the IB from a destructive compression constraint into a constructive coordination process. Cycle formation reduces uncertainty through repeated alignment, while cloning expands the latent space to accommodate structured generalization.
\end{proposition}

\paragraph{Graph-Based Inference: from Loopy Belief Propagation to Chain-based Cognitive Trajectories.}
Beyond its utility in resolving local inference ambiguity, cloning has played a foundational role in extending inference algorithms to more complex structures \cite{ihler2005loopy}. In particular, \emph{cloning has been shown to restore the optimality of loopy belief propagation (LBP)} on arbitrary graphs by transforming cyclic dependencies into conditionally independent copies of variables within an expanded tree-like architecture \cite{weiss2001optimality}. This structural transformation allows inference to remain locally tractable (for context disentangling) while preserving global consistency (for content binding), a principle directly aligned with CCUP's goal of balancing generalization with specificity. 
More recently, the cloning trick has been applied to cognitive modeling: \emph{clone-structured cognitive graph models} explicitly represent overlapping and aliased contextual states as structurally duplicated latent nodes, enabling agents to infer spatial layouts and latent goals even in aliased environments \cite{george2021clone}. 
Importantly, we extend the optimality of LBP from arbitrary graphs \cite{weiss2001optimality} to cognitive trajectories next (a sketch of the proof is referred to Appendix C).

\begin{theorem}[Loopy Belief Propagation Generalizes to Memory Chains under CCUP]
Let $\{Z_1, Z_2, \dots, Z_T\}$ denote a memory chain, where each $Z_t$ is a latent variable representing a memory state embedded in a context-content inference cycle. Suppose the transitions among latent variables form a chain with recurrent loops induced by top-down reconstruction and lateral association.
Under CCUP, the system seeks to minimize joint uncertainty through cycle-consistent inference:
$\min_{q(Z_1, \dots, Z_T)} \sum_{t=1}^{T} \left[ H(Z_t \mid \Psi_t) + H(\Psi_t \mid Z_t) \right]$,
where $q(Z_1, \dots, Z_T)$ is the approximate posterior over the latent trajectory.
Then, iterative inference over this memory chain can be approximated by a loopy belief propagation (LBP) algorithm that minimizes the Bethe free energy:
$\mathcal{F}_{\text{Bethe}} = \sum_{t} H(q_t) - \sum_{(t, t')} I(q_{tt'})$,
where $q_t$ and $q_{tt'}$ are the singleton and pairwise marginals, respectively.
\end{theorem}
\textbf{Remark.}
It follows from Theorem 3 that memory retrieval, imagination, and consolidation emerge as fixed points of message passing that align content and context under cyclic entropy minimization. Thus, LBP generalizes to cognitive memory chains as an emergent mechanism for approximate inference consistent with CCUP. 
Within the CCUP framework, approximate inference is not solely a matter of algorithmic optimization but is fundamentally shaped by how latent representations encode content and context. The quality and structure of these representations determine how efficiently uncertainty can be minimized through iterative message passing \cite{kingma2019introduction}. In particular, inference becomes tractable when representations are initialized in ways that preserve specificity while remaining amenable to contextual refinement. This highlights the critical need for carefully designed priors that can serve as anchors for selective inference \cite{parr2022active}. 
Next, we elaborate on the algorithmic implementation based on predictive coding \cite{keller2018predictive} and introduce a novel delta seeding strategy as implicit representation of context and content variables \cite{ho2022classifierfree}.

\subsection{Latent Space Navigation via Predictive Coding and Delta Seeding}

While cycle formation with cloning provides essential architectural strategies for resolving IB in OT under CCUP, a key question remains: \emph{What neural mechanism enables these operations in a unified and scalable way across the brain?} We propose that \emph{predictive coding} serves as a universal implementation framework \cite{bastos2012canonical}, one that naturally supports both graph-based inference and the contextual cloning of latent structure.
Predictive coding models posit that cortical circuits are organized into hierarchies of generative models, where each level attempts to predict the activity of the level below \cite{friston2005theory}. Crucially, this architecture implements a \emph{recursive inferential cycle}, where beliefs at each level are continuously revised until dynamic consistency is achieved.
From the CCUP perspective, predictive coding performs \emph{contextual disambiguation and content refinement as complementary operations} \cite{rao1999predictive}. Formally, we have the following:

\begin{definition}[Predictive Coding as a Universal Chain-Based Resolution Mechanism]
Under the CCUP framework, predictive coding provides a universal mechanism for resolving the IB in OT by implementing chain-consistent inference across hierarchically cloned latent spaces. Specifically:
1) \textbf{Top-down predictions} encode contextual priors that condition latent content representations at each level of abstraction;
2)\textbf{Bottom-up prediction errors} act as transport gradients that adjust both content estimates and contextual scaffolds;
3) \textbf{Bidirectional inference} generates cycles that progressively reduce joint uncertainty $H(\Phi, \Psi)$, where $\Phi$ denotes content and $\Psi$ denotes context, without requiring simultaneous full specification.

\end{definition}

\textbf{Remark:} Predictive coding resolves the CCUP-induced bottleneck by embedding OT into recursive inference loops, enabling scalable and modular disambiguation of content-context interactions through temporal and hierarchical structure, which we will elaborate next.
These bidirectional updates instantiate cycle formation: each inference pass updates both content and context, gradually aligning latent representations across the hierarchy and reducing directional uncertainty. This process mirrors the refinement of transport plans in OT, but is distributed in time and across layers of abstraction. While predictive coding resolves the context–content bottleneck through chain-based alignment, its inferential machinery also sets the stage for a deeper function: {\bf the simulation of causality} \cite{pearl2009causality}. Once latent content representations are stabilized through recursive inference, they can be reused not just to interpret the past or present, but to simulate counterfactual outcomes and generate possible futures. This transition, from cyclic alignment to generative modeling, reveals predictive coding as more than a perceptual framework; it becomes a generative engine for causal inference. In this view, memory chains that reduce uncertainty also enable structured interventions and “what-if” reasoning, embedding the calculus of causes into the dynamics of inference itself \cite{pearl2018book}.

\paragraph{Memory Chain as a Mechanism for Causal Inference.}
Under the CCUP, a \emph{memory chain} provides a computational framework for dynamic causal inference by embedding counterfactual reasoning within a temporally extended latent structure. Each link in the chain encodes a locally resolved cycle between content and context, while the entire sequence approximates a conditional trajectory through latent space. This trajectory functions analogously to a structural causal model (SCM), where internally generated context shifts or imagined outcomes act as interventions akin to Pearl's do-operator \cite{pearl2009causality}. Through recursive inference, the system simulates alternative possibilities and propagates their consequences forward, enabling \emph{counterfactual evaluation} and goal-driven planning even in the absence of external data \cite{wang2018prefrontal}. Thus, the memory chain supports abduction (hypothesis selection), intervention (goal setting), and prediction (simulation), aligning with Pearl’s triad of causal reasoning \cite{pearl2018book}. This positions memory not merely as storage but as an active mechanism for generating explanations and navigating complex environments through latent causal modeling.

\begin{corollary}[Memory Chain as a Mechanism for Causal Inference]
Under the Context-Content Uncertainty Principle (CCUP), a \emph{memory chain} $\{Z_t\}_{t=1}^T$ forms a sequential latent trajectory in which each state $Z_t$ encodes a partial inference over context $\Psi_t$ and content $\Phi_t$.
This structure supports a dynamic form of causal inference wherein:
1) Each link $Z_t$ represents a locally resolved inference cycle conditioned on prior latent variables, analogous to abduction in Pearl’s causal inference;
2) Internally generated goals or modifications to context-content priors act as \emph{interventions}, similar to Pearl’s $\text{do}(X = x)$ operator, altering the flow of inference within the chain;
3) The forward simulation of consequences and backward revision of causes along the chain approximates counterfactual reasoning, enabling the system to evaluate alternative actions, simulate hypothetical scenarios, and refine decision policies.

\end{corollary}

\textbf{Remark:} 
The memory chain serves as an internalized mechanism for counterfactual simulation and causal explanation, unifying goal-directed planning and structural reasoning within a generative latent framework.
While memory chains implement temporal bootstrapping by aligning latent representations across sequential inference cycles, cognitive systems must also resolve uncertainty across spatial or hierarchical scales. This necessitates a complementary form of bootstrapping, \emph{spatial bootstrapping}, in which content and context are recursively disentangled and bound across abstraction levels. We propose that this is achieved through \emph{delta-seeded anchors} \cite{ho2022classifierfree}, wherein inference at each hierarchical level is initialized with sharply concentrated (delta-like) latent priors that selectively propagate uncertainty through structured diffusion. Unlike temporal bootstrapping, which relies on recurrence and context accumulation over time \cite{song2021score}, spatial bootstrapping organizes inference vertically, such that lower levels contribute fine-grained context while higher levels impose top-down constraints on content binding. By integrating both forms of bootstrapping, the system achieves dynamic consistency not only across time but also across abstraction, allowing latent representations to stabilize through bidirectional refinement at multiple spatial and temporal scales.

\section{Hierarchical Representation of Content and Context via Delta-Seeded Diffusion Models}
\label{sec:4}


To operationalize CCUP in complex systems, we propose a novel mechanism: \emph{delta-seeded latent diffusion}. This approach formalizes how systems can encode/bind content and decode/disentangle context through selective inference processes distributed across multiple levels of abstraction \cite{friston2008hierarchical}. In particular, we show that delta-seeded diffusion provides a generative and computational framework for implementing cycle-consistent inference under CCUP at each level of a hierarchical architecture. Moreover, predictive coding naturally supports the implementation of \emph{hierarchically cloned latent representations} by canonical circuits \cite{keller2018predictive}. At each level, latent variables can be specialized into context-specific variants of shared content, enabling independent inference cycles that remain globally coherent. For example, a generic ``face'' template represented by a delta seed (e.g., polychronization neural group \cite{izhikevich2006polychronization}) may be cloned into several variants across lighting, distance, or pose. This prevents aliasing between similar structures encountered in different contexts, ensuring that the latent space remains navigable and distinct \cite{bastos2012canonical, millidge2021predictive}.

\subsection{Delta-Seeding Anchors Content Binding and Context Disentangling}

In conventional diffusion models \cite{ho2020denoising}, sampling is initiated from noise, and structure emerges through a generative denoising process. In contrast, \emph{delta-seeded diffusion} initiates inference from an idealized, sharply localized representation, a delta function in latent space \cite{ballard1981generalizing}. This delta-seed acts as an anchor for content, enforcing a low-entropy prior and high precision in representational space. The diffusion process then allows contextual uncertainty to unfold around this anchor, forming a distribution that represents plausible contextual variations.
Formally, let $X \in \Psi$ be a sensory observation representing a contextual variable and $z_0 \in \Phi$ denote the delta-seed, representing a hypothesized content state. A latent diffusion process generates context-conditioned representations $z_t$ over time $t$ ($\theta$ is the parameter in the latent space):$z_t = \mathcal{D}(z_0, t; \theta),$
where $\mathcal{D}$ is a parametric denoising or stochastic evolution function. The delta-seed ensures strong content binding by minimizing $H(Z_0)$, while the unfolding distribution models context variability $H(X \mid Z)$ following CCUP. This asymmetric evolution mirrors the trade-off between content precision and context richness. A good example illustrating such content-context tradeoff is the generalized Hough transform (GHT) for shape recognition \cite{ballard1981generalizing}.

\begin{tcolorbox}[title=\textbf{Anchor Example: Delta-Seeding in Generalized Hough Transform (GHT)}, colback=gray!5, colframe=blue, fonttitle=\bfseries]
The Generalized Hough Transform (GHT) offers a classical yet powerful example of how \emph{delta-seeded inference} anchors content binding and facilitates context disentangling. In GHT \cite{ballard1981generalizing}, aggregation collects evidence from local features by mapping them into a common parameter space, while voting accumulates this evidence across instances to identify the most likely global configuration of the target object.
Equivalently, a specific feature anchor, such as a closed shape in a 2D plane (the composition of multiple line segments), is selected as a \textbf{delta seed}: a precise, low-entropy content descriptor that defines the object’s identity (i.e., shape). Local features (e.g., edges or gradients) are then mapped relative to this seed in a latent accumulator space, where evidence from distributed observations is integrated to vote for object presence regardless of contextual variations.

\begin{itemize}
  \item \textbf{Content Binding:} The delta seed binds local features into a coherent object hypothesis, defining a stable content representation across transformations.
  \item \textbf{Context Disentangling:} By aligning observed features to the reference seed, GHT disentangles object identity from contextual variations such as translation, rotation, and scale.
\end{itemize}

This mirrors the role of delta-seeded diffusion in cycle formation under CCUP framework: inference begins from an aliased observation $X$ and encoding reaches a precise latent anchor $Z$, the cyclic diffusion selectively unfolds across contextual dimensions to reconstruct $\hat{X}$ and disambiguate the surrounding structure $Z'$.
\end{tcolorbox}

\subsection{Hierarchical CCUP for Joint Content Binding and Context Disentangling}

The challenge of content binding and context disentangling becomes especially acute in hierarchical systems, where features at each layer must be integrated into coherent representations while discarding irrelevant variability. We extend CCUP to hierarchical architectures by positing that each layer alternates between two complementary roles:
\begin{itemize}
    \item \textbf{Bottom-up selective contextualization}: extracting latent features $Z^{(l)}$ from ambiguous inputs $X^{(l)}$, conditioned on context from the level below.
    \item \textbf{Top-down manifold-constrained reconstruction}: disambiguating $Z^{(l)}$ by enforcing consistency with higher-level abstractions $Z^{(l+1)}$.
\end{itemize}
This leads to a two-way inference mechanism per layer that mirrors CCUP: bottom-up processes encode and bind content variables, while top-down processes condition and disentangle contextual variables. The hierarchical structure ensures that contextual variations at lower levels are systematically filtered out by predictive constraints from above, while content hypotheses are refined by recurrent cycles. Next, we extend the design principles of bidirectional selective inference into hierarchical systems.

\paragraph{Hierarchical Extension of Dynamic Alignment.}
The principle of dynamic alignment through context disambiguation and reconstruction naturally extends across hierarchical levels of cognitive inference \cite{friston2008hierarchical}. In hierarchical systems, whether biological or artificial, lower layers perform rapid, fine-grained disambiguation of sensory input, while higher layers accumulate more abstract, temporally extended contextual information. Under CCUP, each level of the hierarchy engages in its local inference cycle, aligning content and context through bottom-up conditioning and top-down reconstruction \cite{kingma2014auto}. Crucially, these cycles are not isolated but are recursively coupled: the disambiguated content from one level serves as contextual input for the next, and higher-level abstractions guide lower-level predictions \cite{dayan1995helmholtz}. This nested structure allows uncertainty to be progressively redistributed across layers, enabling both specificity at the sensory periphery and generalization at abstract levels. The result is a scalable architecture in which inference converges not only within each level but also across the hierarchy, forming a cascade of dynamically aligned cycles. Such multiscale coordination explains how complex cognitive systems integrate real-time perception, memory, and abstraction within a unified inferential framework \cite{clark2013whatever}. Formally, we can extend Proposition 1 as follows.

\begin{proposition}[Hierarchical Cycle Completion under CCUP with Delta-Seeding]
Let a hierarchical system be composed of \( L \) levels of abstraction, each with latent representations \( Z^{(1)}, \dots, Z^{(L)}=\Phi \), and contextual observations \( \Psi = Z^{(0)} \). Under CCUP, each level \( \ell \in \{1, \dots, L\} \) performs a local inference cycle:
$Z^{(\ell - 1)} \xrightarrow{\text{encoding}} Z^{(\ell)} \xrightarrow{\text{delta-seeded diffusion}} \hat{Z}^{(\ell - 1)} \xrightarrow{\text{update}} Z^{(\ell)\prime}$
such that:
\begin{enumerate}
    \item \textbf{Bottom-up binding:} \( Z^{(\ell)} \) is initialized by an anchor seed (e.g., GHT) as a low-entropy delta distribution \( \delta(z^{(\ell)}_0 \mid Z^{(\ell - 1)}) \), enforcing strong content anchoring.
    \item \textbf{Top-down disentangling:} A constrained diffusion process \( \mathcal{D}^{(\ell)} \) unfolds \( z^{(\ell)}_0 \) into context-conditioned predictions: $\hat{Z}^{(\ell - 1)} = \mathcal{D}^{(\ell)}(z^{(\ell)}_0, t; Z^{(\ell + 1)})$   
    guided by higher-level constraints \( Z^{(\ell + 1)} \).
    \item \textbf{Cycle-consistent feedback:} Discrepancy between \( Z^{(\ell - 1)} \) and \( \hat{Z}^{(\ell - 1)} \) is measured:
    $\delta^{(\ell)} = d\left(Z^{(\ell - 1)}, \hat{Z}^{(\ell - 1)}\right),$
    and used to refine \( Z^{(\ell)} \) through a self-supervised update:
    $Z^{(\ell)\prime} = Z^{(\ell)} - \eta \nabla_{Z^{(\ell)}} \delta^{(\ell)}.$
\end{enumerate}

If each local cycle minimizes \( \delta^{(\ell)} \leq \epsilon \), then the full system converges toward a globally cycle-consistent hierarchy, minimizing conditional entropy at each level and maximizing mutual information:
$I(Z^{(\ell - 1)}; Z^{(\ell)}) \approx H(Z^{(\ell - 1)}).$
This resolves the IB across levels by dynamically aligning content and context representations under delta-seeded inference.
\end{proposition}

\subsection{Bidirectional Selective Inference via Delta-Seeded Diffusion Across Levels}

At each level $l$ in the hierarchy, bidirectional selective inference is implemented via delta-seeded diffusion:
\begin{equation}
z^{(l)}_t = \mathcal{D}^{(l)}\left(\delta\big(z^{(l)}_0 \mid Z^{(l-1)}\big), t; Z^{(l+1)}\right),
\end{equation}
where $Z^{(l-1)}$ provides the bottom-up contextual prior used to initialize the seed $z^{(l)}_0$ and $Z^{(l+1)}$ provides top-down constraints that shape the evolution of the diffusion process $\mathcal{D}^{(l)}$.
This setup allows each level to:
1) bind features into a stable representation by anchoring to a sharp initial seed (e.g., place cells \cite{moser2008place} or concept cells \cite{quiroga2005invariant}); and
2) disentangle irrelevant contextual variation via guided evolution constrained by higher-level structures (e.g., hippocampal remapping \cite{colgin2008understanding}).
Thus, inference proceeds not as a single feedforward computation, but as a cyclic, self-refining process across the hierarchy, where each level aligns content with relevant context while reducing ambiguity from irrelevant signals.
By integrating delta-seeded latent diffusion with hierarchical CCUP, we propose a biologically plausible and computationally powerful mechanism for joint content binding and context disentangling. This framework explains how complex systems, from mammalian brains \cite{bennett2023brief} to generative models \cite{rombach2022high}, can maintain structured, interpretable representations under high-dimensional uncertainty while preserving information flow through cycles at every level of abstraction.
However, disentangling context (from content) tends to require more metabolic and computational effort than binding content (from context), because it involves constructive inference over ambiguous high-dimensional variables (please refer to the sidebar below on broken symmetry). Next, we elaborate on how CCUP overcomes the barrier with the curse of dimensionality by starting with low-entropy latent content (delta seeding), and diffusing outward into the high-dimensional observation space, guided by a structured generative model \cite{song2021score}.

\paragraph{Manifold-constrained Context Reconstruction.}
Top-down context reconstruction refers to the generative process of reconstructing plausible contextual configurations from compact content representations. However, rather than exploring the full high-dimensional space of possible contexts, this reconstruction is constrained to unfold along a learned low-dimensional manifold shaped by prior experience and structural regularities. These manifolds encode the lawful variations of context conditioned on latent content, such as typical spatial arrangements, temporal sequences, or semantic associations, thereby guiding reconstruction toward coherent and generalizable interpretations \cite{higgins2017beta}. Within predictive coding architectures, this constraint is implemented by top-down predictions that are not arbitrary but restricted to lie on trajectories supported by generative priors. This manifold constraint serves two critical functions: 1) it regularizes inference by preventing overfitting to unstructured contextual variations; and 2) it promotes abstraction by aligning new observations with previously learned contextual patterns. In this way, manifold-constrained reconstruction ensures that the inverse mapping from content to context remains both efficient and meaningful, enabling robust generalization and the contextual reuse of compact representations. Formally, we have

\begin{corollary}[Delta-Seeded Diffusion Overcomes the Curse of Dimensionality]
Let $\delta(z_0)$ be a delta measure over a low-dimensional latent variable $z_0 \in \mathcal{Z}$, and let $p(x \mid z)$ denote a generative model mapping content $z$ to observation $x \in \mathcal{X}$, where $H(\mathcal{X}) \gg H(\mathcal{Z})$. Under CCUP, if the generative mapping defines a differentiable manifold $\mathcal{M} \subset \mathcal{X}$ of bounded intrinsic dimension, then:

\begin{enumerate}
    \item Delta-seeded diffusion $\tilde{x}_t \sim q_t(x \mid z_0)$ approximates samples from $p(x \mid z_0)$ along the manifold $\mathcal{M}$, avoiding the need to explore the full ambient space $\mathcal{X}$.
    
    \item The effective sampling complexity grows with the intrinsic dimension of $\mathcal{M}$, not with $\dim(\mathcal{X})$, thereby avoiding the curse of dimensionality.
    
    \item When integrated into a hierarchical cycle, where each level conditions on the delta-seeded content from below and constraints from above, the total inference complexity scales additively across levels rather than exponentially, enabling efficient selective inference.
\end{enumerate}
\end{corollary}

\noindent \textbf{Remark.} It follows from Corollary 2 that delta-seeded diffusion enables tractable alignment between high-entropy context and low-entropy content by confining uncertainty expansion to context-relevant directions on $\mathcal{M}$. Delta-seeded diffusion generalizes and formalizes the grandmother cell idea \cite{quiroga2005invariant} by framing these sparse, content-specific representations as seeds that dynamically generate and disentangle broader contextual information through a diffusion-like unfolding process in latent space. Moreover, delta-seeded diffusion across scales calls for a scalable extension of resolving IB for OT in hierarchical systems.


A complementary intuition for resolving the IB in a scalable OT solution arises from the theory of hierarchical navigable small-world (HNSW) networks \cite{malkov2018efficient}. These networks enable efficient search and routing by embedding latent representations across nested spatial or semantic scales. Under the CCUP framework, the asymmetry of directional information loss implies that the IB reappears recursively: disambiguating local context while preserving global structure requires resolving uncertainty not just once, but at every scale of inference \cite{friston2008hierarchical}. To address this, a \emph{hierarchical extension of the cloning trick} is required. At each level of abstraction, cloned latent representations can capture context-specific variants of globally relevant content, allowing inference cycles to operate independently yet coherently across the hierarchy \cite{bengio2013representation, esmaeili2019structured}. This hierarchical cloning ensures that latent structure remains navigable and distinct, preventing aliasing between similar substructures encountered at different scales. As a result, the agent's internal representation mirrors the nested, compositional structure of the external world. In this view, hierarchical cloning enables scalable generalization without sacrificing specificity, and supports efficient inference and goal-directed behavior across the full spectrum of context, from local sensory details to abstract task space.

\begin{proposition}[Fractal-like HNSW Inference Resolves IB in Scalable OT]
Let $\mathcal{X}$ denote the high-entropic observation space, and $\mathcal{Z}$ a lower-entropic latent content space. Let $T: \mathcal{Z} \to \mathcal{X}$ be a generative transport map. Under the CCUP, the IB in scalable OT can be resolved by embedding inference within a fractal-like HNSW architecture satisfying the following:
1) \textbf{Delta-seeded diffusion}: Inference is initiated from a delta seed $z_0 \in \mathcal{Z}_L$ (high-level content), and proceeds downward through structured diffusion constrained by top-down priors and bottom-up sensory context, forming a selective content-context cycle;
2) \textbf{HNSW-style connectivity}: Each level $\mathcal{Z}_\ell$ is embedded in a small-world graph with both local connections (for fine-grained refinement) and sparse long-range links (for global coherence), enabling efficient approximation of transport paths with sublinear complexity.   
\end{proposition}

\textbf{Remark.} Under the above conditions, the inference process avoids the curse of dimensionality by operating on the intrinsic dimension of $\mathcal{M} \subset \mathcal{X}$ while maximizing mutual information $I(X;Z)$ by cycle-consistent transport across layers and resolving the IB in OT by enabling selective, hierarchical inference from low-entropy content to high-entropy context. The hierarchical organization of the brain closely resembles an HNSW \cite{malkov2018efficient}: each cortical layer acts as a locally clustered module that encodes partial, low-dimensional views of the world, while long-range interconnections enable efficient transitions across distant representational spaces. Just as HNSW performs sublinear approximate nearest neighbor search by navigating from coarse-grained to fine-grained levels with shortcut links, the neocortex organizes inference across multiple abstraction scales, starting from low-entropy latent seeds \cite{barlow1972single} and gradually expanding outward through nested representational manifolds. This design enables the brain to transport information efficiently through high-dimensional latent spaces without suffering from a combinatorial explosion. Instead of compressing content representations into a narrow bottleneck, the cortical hierarchy distributes inference across levels in a manner that maximizes information flow and minimizes entropy misalignment between context and content. 

\begin{tcolorbox}[title={\textbf{Broken Symmetry: Energetic Asymmetry in Inference Under CCUP}}, colback=gray!5, colframe=blue, fonttitle=\bfseries]

\textbf{CCUP posits a fundamental asymmetry between two core inference operations:}

\vspace{1mm}
\begin{itemize}
  \item \textbf{Content Binding} (\( \Psi \rightarrow \Phi \)) condenses high-entropy context into stable, low-entropy latent content. This is typically fast, sparse, and energetically efficient, analogous to pattern recognition or memory recall.
  \item \textbf{Context Disentangling} (\( \Phi \rightarrow \Psi \)) reconstructs or simulates diverse contextual instances from compact latent seeds. This process is expansive, generative, and metabolically costly, analogous to imagination, dreaming, or counterfactual reasoning.
\end{itemize}

\vspace{2mm}
\noindent\textbf{Energetic Implication:} 
$\text{Energy}(\Phi \rightarrow \Psi) \gg \text{Energy}(\Psi \rightarrow \Phi)$
This reflects the higher computational demands of exploring ambiguous, high-dimensional context spaces during generative reconstruction. Empirical evidence from sleep and dreaming (e.g., lucid dreaming \cite{baird2019cognitive}) supports this asymmetry. Under CCUP, inference is not only directionally asymmetric in terms of entropy but also in energetic cost. This has implications for attention, learning, and the design of memory architectures in both biological and artificial systems.

\end{tcolorbox}

\subsection{Chain Formation as Cognitive Renormalization}

Under the CCUP framework, chain formation arises from the iterative composition of content–context inference cycles into temporally extended latent trajectories. Each cycle locally minimizes conditional uncertainty via bidirectional message passing between latent content $\Phi_t$ and context $\Psi_t$. When such cycles are embedded into a structured chain and approximated via LBP \cite{weiss2001optimality}, the system constructs progressively refined beliefs along a trajectory that supports memory, imagination, and goal-directed behavior.
This structured inference process parallels the logic of the Renormalization Group (RG) in statistical physics \cite{wilson1974renormalization}, which describes how system parameters evolve under coarse-graining across spatial or energetic scales. In RG theory, transformations are applied to microscopic configurations to obtain effective macroscopic descriptions while preserving relevant informational content. Similarly, in chain-based LBP, inference proceeds across levels of abstraction by integrating fine-grained bottom-up evidence with top-down constraints, effectively performing a scale-dependent belief refinement.
Recent work has highlighted formal connections between inference and RG - e.g., \cite{mehta2014exact} showed that deep learning architectures perform transformations analogous to RG flows and \cite{beny2013quantum} demonstrated that quantum belief propagation aligns with information-preserving RG operations. The IB method, itself a core part of CCUP, has also been interpreted as an RG-like framework that compresses irrelevant degrees of freedom while preserving predictive relevance \cite{tishby2011information}.
Under this view, each segment of a memory chain acts as a local renormalization step: belief states evolve under contextually constrained inference dynamics, and global consistency is achieved through alignment across the entire chain. When delta-seeded diffusion is applied hierarchically, spatial bootstrapping complements the temporal chain, together forming a bi-directional renormalization structure across time and abstraction \cite{friston2017active}. Formally, we have the following main theorem.

\begin{theorem}[Cognitive Renormalization via Chain-Based Inference]
Let $\{(\Psi_t, \Phi_t)\}_{t=1}^T$ be a memory chain under the CCUP framework, where each cycle minimizes local uncertainty through structured Loopy Belief Propagation. Suppose each step propagates marginal beliefs forward and backward using Rao-Blackwellized messages conditioned on delta-seeded latent priors. Then the entire chain induces an information-preserving flow over latent variables:
$q(Z_t) \mapsto q(Z_{t+1}) \quad \text{with} \quad I(Z_t; Z_{t+1}) \approx \max$,
such that the system constructs increasingly abstract yet coherent representations across time. This inference trajectory is equivalent to a renormalization flow that minimizes irrelevant entropy while maintaining predictive and goal-relevant structure. Thus, chain formation serves as a cognitive implementation of renormalization, dynamically aligning latent representations across both spatial and temporal scales.
\end{theorem}

\textbf{Remark.} A sketch of the proof can be found in Appendix D. 
Having established cognitive renormalization as a structured inference process analogous to renormalization group flows, we now turn our attention to the computational implementation of this framework. Traditional inference methods on structured graphical models, such as Markov Random Fields (MRFs) \cite{geman1984stochastic}, face significant computational challenges due to the necessity of evaluating global Gibbs potentials. To address this limitation, we propose the adoption of a delta-seeded diffusion model, a strategy that dramatically simplifies inference by initializing latent states with sharply localized distributions. In what follows, we demonstrate how this method not only aligns conceptually with the hierarchical inference of cognitive renormalization but also achieves substantial computational efficiency by eliminating explicit calculation of the Gibbs potentials characteristic of conventional inference algorithms.


\paragraph{Computational Efficiency of Delta-Seeded Diffusion.}
Chain-based cognitive renormalization via structured Loopy Belief Propagation (LBP) naturally corresponds to inference in Markov Random Fields (MRFs) \cite{geman1984stochastic}. Consider a latent memory chain $\{Z_t\}$, where each latent state comprises context ($\Psi_t$) and content ($\Phi_t$). In an MRF formulation, these latent states form nodes, and their conditional dependencies constitute the edges of the graphical model \cite{wainwright2008graphical}. Here, local potentials correspond directly to the immediate context variables, reflecting sensory inputs or temporally localized priors, while global potentials represent content-level constraints—stable, long-term dependencies derived from experience or learned hierarchical structures \cite{friston2017graphical, dayan1995helmholtz}.
Traditionally, inference in MRFs requires computation of a Gibbs distribution:
$p(Z) = \frac{1}{\mathcal{Z}} \exp\left(-E(Z)\right),\quad\mathcal{Z} = \sum_Z \exp(-E(Z))$,
where the normalization constant $\mathcal{Z}$ (partition function) is computationally intractable for large-scale latent spaces \cite{koller2009probabilistic}. This computational bottleneck makes exact inference prohibitively expensive, necessitating approximate methods.

In contrast, \emph{delta-seeded diffusion} achieves substantial computational efficiency by avoiding explicit computation of these expensive Gibbs potentials. By initializing inference with sharply peaked, delta-like priors, the diffusion model converts the inference process into a sequence of local deterministic or semi-deterministic propagation steps:
$q(Z_t \mid Z_{t-1}) \approx \delta(Z_t - f(Z_{t-1})) + \epsilon_t$,
where $f(\cdot)$ is a deterministic or near-deterministic function and $\epsilon_t$ represents local uncertainty. This formulation obviates the need for global integrals or partition function evaluations, leveraging highly localized conditional updates. Consequently, delta-seeded diffusion supports scalable and biologically plausible inference by significantly reducing computational complexity.

\begin{corollary}[Computational Efficiency via Delta-Seeded Diffusion]
Let $\{Z_t\}_{t=1}^{T}$ form a latent memory chain under cognitive renormalization structured via LBP. Suppose the local potentials correspond to context variables and global potentials correspond to content variables within an MRF formulation. If inference is implemented via delta-seeded diffusion, then the computational complexity associated with explicit evaluation of Gibbs potentials is eliminated, and inference reduces to efficient local propagation of sharply peaked conditional distributions. Thus, delta-seeded diffusion represents an efficient, scalable cognitive inference strategy aligned with both CCUP and biological plausibility.
\end{corollary}

\textbf{Remark.}
Under a Gibbs distribution, inference demands costly global normalization through the partition function $\mathcal{Z}$. Delta-seeded diffusion sidesteps this computation by initializing latent states using sharply peaked (delta-like) priors. Consequently, double bootstrapping of inference (see sidebar below) updates at each step become local deterministic or near-deterministic propagations without extensive marginalization or normalization integrals. Thus, the computational cost of bidirectional selective inference reduces from exponential or high-dimensional integral evaluations to linear, locally deterministic message-passing operations.

\begin{tcolorbox}[title=Dual Bootstrapping of Inference: Temporal Chains and Spatial Hierarchies, colback=gray!5, colframe=blue, width=\textwidth, boxrule=0.5pt]

\begin{center}
\begin{tikzpicture}[node distance=1.6cm, every node/.style={align=center}, scale=1, transform shape]

\node (psi1) [draw, circle] at (0,0) {$\Psi_1$};
\node (phi1) [draw, circle, right of=psi1, xshift=1.5cm] {$\Phi_1$};
\node (psi2) [draw, circle, right of=phi1, xshift=1.5cm] {$\Psi_2$};
\node (phi2) [draw, circle, right of=psi2, xshift=1.5cm] {$\Phi_2$};
\node (dots) [right of=phi2, xshift=1.5cm] {$\cdots$};

\draw[->] (psi1) -- (phi1);
\draw[->] (phi1) -- (psi2);
\draw[->] (psi2) -- (phi2);
\draw[->] (phi2) -- (dots);

\node at (5, -2.0) {\small \textbf{Temporal Bootstrapping (Memory Chain)}};

\draw[->, thick, dashed] (psi1) -- +(0,1.3) node[midway, left] {\tiny $\uparrow$};
\draw[->, thick, dashed] (phi1) -- +(0,-1.3) node[midway, right] {\tiny $\downarrow$};
\draw[->, thick, dashed] (psi2) -- +(0,1.3);
\draw[->, thick, dashed] (phi2) -- +(0,-1.3);

\node at (0, 1.5) {\small Higher-level Context};
\node at (3.0, -1.5) {\small Lower-level Content};

\node at (5, 2.0) {\small \textbf{Spatial Bootstrapping (Delta-Seeded Diffusion)}};

\end{tikzpicture}
\end{center}

\vspace{0.5em}
\noindent\small
\textbf{Summary:} Inference in the brain bootstraps representations along two complementary dimensions. \emph{Temporal bootstrapping} occurs via memory chains—sequential context-content inference cycles that align representations over time. \emph{Spatial bootstrapping} unfolds through hierarchical inference: each level diffuses uncertainty conditioned on sharp delta-like seeds from below and constrained by contextual priors from above. Together, these dual processes ensure both stability and generalization in cognitive computation.

\end{tcolorbox}

\section{Applications: from Ventral-Dorsal Separation to the Origin of Language}
\label{sec:5}

\subsection{Object Recognition: Ventral-Dorsal Separation}
\label{sec:5.1}

This spatial separation of the “what” and “where/how” pathways \cite{ungerleider2000mechanisms} is deeply consistent with the CCUP framework, which posits an intrinsic asymmetry between low-entropy content representations and high-entropy contextual structures. The ventral stream specializes in refining object-specific content $\Phi$ through local feature integration and transformation-invariant recognition \cite{dicarlo2007untangling}. It achieves this by projecting sensory input into a relatively low-dimensional, stable latent subspace that is ideal for content binding \cite{treisman1980feature}. In contrast, the dorsal stream operates over broader spatial and motor-contingent variables such as egocentric position, object affordances, and task-relevant goals, representing high-entropy context $\Psi$ that must be disambiguated over time and space. The separation is therefore not modular in the traditional sense but functional: each stream exploits the entropy asymmetry in a complementary fashion to reduce uncertainty during bidirectional selective inference \cite{goodale1992separate}.

Moreover, this division supports a critical augmentation of the ventral stream’s function in object recognition. While the ventral stream performs local “untangling” of object features into separable manifolds (as emphasized in DiCarlo’s hypothesis on cortical representation \cite{dicarlo2012does}), this untangling is heavily scaffolded by the dorsal stream’s top-down priors. These priors encode contextual constraints such as spatial layout, attention allocation, and task-driven expectations, effectively conditioning the ventral stream’s representational subspace. This top-down modulation helps resolve feature ambiguity, especially under occlusion, clutter, or uncertainty. Crucially, the dorsal stream does not merely guide “where” to look—it also supplies abstract context that is simultaneously encoded with object identity, thereby reducing aliasing and improving generalization \cite{jeannerod1994representing}. In this way, dorsal priors enable selective gating and amplification of ventral hypotheses, allowing the system to form coherent object representations that are grounded in scene context.

From a CCUP standpoint, this interaction solves the joint problem of content binding and context disentangling through \emph{spatial bootstrapping}. Content binding is achieved by aggregating ventral features into stable object codes conditioned on dorsal predictions, while context disentangling is enabled by interpreting those object codes relative to broader scene structure. The iterative exchange of predictions and updates between streams reduces both 
$H(\Psi \mid \Phi)$ and $H(\Phi \mid \Psi)$, forming a bidirectional cycle that aligns object identity with scene context. This spatially structured inference not only improves recognition but also supports sensorimotor control, memory indexing, and semantic abstraction. Ultimately, ventral-dorsal coordination under CCUP transforms the perception of a complex, ambiguous scene into a set of coherent, actionable percepts via cycle-consistent reduction of uncertainty across complementary representational dimensions.


Concurrent encoding of content and contextual variables facilitates joint inference, improves generalization, and reduces spurious feature binding. In effect, the dorsal stream enables the ventral stream to perform inference over structured manifolds, not merely flat feature spaces \cite{olshausen2005toward}.
This leads to a \emph{spatial inference cycle} in which the dorsal stream directs attention and expectation, while ventral representations update and refine those contextual predictions \cite{mante2013context}. The two streams exchange information iteratively to minimize both conditional entropies:
$H(\Phi \mid \Psi) \downarrow \quad \text{and} \quad H(\Psi \mid \Phi) \downarrow$, thus recovering mutual information \( I(\Psi; \Phi) \) across spatial scales. From the CCUP perspective, this cyclical coordination transforms the IB into a scaffold for adaptive perception and action. Our analysis is summarized below.

\begin{observation}[Spatial Bootstrapping Resolves the IB via Ventral–Dorsal Cycles]
Under CCUP, the ventral stream binds features into stable content representations \( \Phi \), while the dorsal stream scaffolds inference by encoding high-entropy contextual variables \( \Psi \). Their bidirectional interaction reduces both \( H(\Phi \mid \Psi) \) and \( H(\Psi \mid \Phi) \), resolving the information bottleneck (IB) not through compression, but through cycle-consistent alignment. This \textbf{spatial bootstrapping} enables robust perceptual inference, generalization, and content–context disentangling across hierarchical representations.
\end{observation}

\subsection{Sensorimotor Learning: Perception–Action Cycle}
\label{sec:4.2}

Sensorimotor learning lies at the heart of intelligent behavior, enabling agents to adaptively interact with their environment by mapping sensory inputs to motor outputs through experience \cite{wolpert1995internal}. From a CCUP perspective, the perception–action cycle (PAC) provides a core inferential mechanism that resolves uncertainty not only by interpreting observations but also by actively shaping them \cite{wolpert2011principles}. A key open question in PAC is: how can intelligent systems dynamically align perception and action representations to enable generalization and transfer across novel tasks and environments? We shed some new insight into this question under the CCUP framework.

\paragraph{The Duality of Perception and Action Under CCUP.}  
Under the CCUP, perception and action emerge as informationally dual yet complementary processes within intelligent systems. Perception operates in a \emph{bottom-up} manner, transforming ambiguous, high-entropy sensory context \( \Psi_s \) into low-entropy, specific latent content \( \Phi_s \), such as object identities, spatial layouts, or scene structure. This process corresponds to \emph{contextual disambiguation}, where the system reduces uncertainty by conditioning on sensory inputs and refining hypotheses about the environment \cite{bar2004visual}.
Conversely, action unfolds in a \emph{top-down} direction: compact motor intentions or plans \( \Phi_a \) serve as low-entropy seeds that generate or reconstruct high-entropy contextual effects \( \Psi_a \) in the external world. Rather than merely executing commands, actions actively \emph{probe} and \emph{modify} the environment to elicit informative sensory consequences, thereby expanding or shaping the context for subsequent perceptual inference. This top-down reconstruction aligns with the CCUP’s principle that inference involves \emph{transforming} content into context, enabling agents to influence and learn from their surroundings \cite{friston2015active1}.
This intrinsic directional asymmetry between perception and action—bottom-up disambiguation versus top-down contextual generation—is fundamental to CCUP \cite{poggio2004generalization}. Crucially, it implies that perception and action cannot be treated as isolated modules but must be tightly coupled through a \emph{cyclic transformation}. This cycle is facilitated by \emph{delta-seeded anchoring}, where precise low-entropy content representations serve as seeds to unfold or constrain the high-dimensional contextual space dynamically. Through this mechanism, perception and action align within a shared latent space, enabling coherent sensorimotor integration and adaptive behavior \cite{sohn2021network}.

\paragraph{Closed-Loop Inference Through Embodied Interaction.}
Crucially, perception and action form a closed loop in embodied agents \cite{shapiro2019embodied}. Rather than functioning as separate modules, perception and action operate as coupled inference processes: perception informs action selection by disambiguating latent variables from noisy sensory input, while action reshapes the agent’s sensory context by modifying the external environment. This reciprocal relationship forms the foundation of \emph{embodied inference}, wherein the agent actively refines its internal generative model through cycles of environmental interaction.
In this loop, action is not merely a reactive motor command—it serves as an \emph{epistemic operator} \cite{friston2010action}. By strategically probing the environment (e.g., shifting gaze, manipulating objects, changing locomotion), action generates novel observations that resolve ambiguity in perceptual inference. This active sampling reduces the conditional entropy \( H(\Phi_s \mid \Psi_s) \) of perceptual content given context and simultaneously constrains the space of plausible contextual reconstructions \( H(\Psi_a \mid \Phi_a) \). 
From the CCUP perspective, this process transforms action from a goal-directed output into a tool for uncertainty minimization. The agent learns not only \emph{what to do} but also \emph{what to sense}, tuning its interaction strategy to maximize epistemic gain \cite{gibson1979ecological}. Thus, the perception–action cycle becomes a self-supervising engine: each iteration of embodied interaction refines both perceptual and motor content variables, enabling more accurate and generalizable inferences over time \cite{pezzulo2014internally}.

\paragraph{Latent Coordination of Content Variables.}  
The effectiveness of the perception–action cycle critically hinges on the precise alignment of latent content variables across sensory and motor domains \cite{sohn2021network}. Specifically, the content inferred through perception, denoted \( \Phi_s \), must be consistently and coherently mapped to the content guiding action, denoted \( \Phi_a \). When these latent representations “click” into a shared space, the agent forms a unified internal model that simultaneously supports robust recognition of environmental features and effective manipulation or control of those features.
This shared latent space acts as a common substrate for sensorimotor integration, enabling information from perception to directly inform action plans, and conversely, for anticipated actions to shape perceptual expectations. Such coordination allows the agent to leverage internal predictions and feedback loops, facilitating rapid adaptation to changing contexts and goals.
Conversely, misalignment or incoherence between \( \Phi_s \) and \( \Phi_a \) leads to a breakdown in this integration, resulting in errors in both inference and motor execution. For example, perceptual ambiguities that fail to align with motor goals can cause inappropriate or ineffective actions, while motor commands not grounded in accurate perceptual content may degrade sensory prediction and learning. Therefore, ensuring latent coordination under CCUP is essential for maintaining the fidelity and utility of internal models, supporting both generalization across tasks and robustness in sensorimotor control.

\paragraph{Maximizing Mutual Information in Sensorimotor Learning.}  
The perception–action cycle (PAC) can be viewed as a continuous, dynamic process aimed at maximizing the mutual information \( I(\Phi_s; \Phi_a) \) between perceptual content variables \( \Phi_s \) and motor content variables \( \Phi_a \) \cite{linsker1990perceptual}. By iteratively refining motor policies informed by perceptual predictions—and simultaneously updating perceptual inference based on motor outcomes—the agent cultivates sensorimotor mappings that effectively reduce uncertainty about both sensory inputs and motor consequences.
This reciprocal refinement aligns with the CCUP-driven framework of cycle formation, wherein the agent minimizes the conditional entropies \( H(\Phi_s \mid \Psi_s) \), representing uncertainty in perceptual content given sensory context, and \( H(\Psi_a \mid \Phi_a) \), representing uncertainty in the environmental context generated by action. Joint minimization of these entropies enhances the fidelity and practical utility of the agent’s internal representations across both perception and action modalities.
Consequently, sensorimotor learning emerges as a form of embodied active learning: the agent’s intelligent behavior is not the product of passive observation or unguided trial-and-error but results from context-sensitive, content-aligned interaction cycles that strategically sample and adapt to the environment \cite{pezzulo2014internally}. This perspective highlights how information-theoretic principles underpin the tight coupling between perception and action, supporting robust adaptation and generalization in complex, dynamic settings.

\begin{observation}[Sensorimotor Learning as Cycle-Consistent Mutual Alignment]
The perception–action cycle implements a CCUP-guided mechanism whereby perception reduces uncertainty through bottom-up contextual disambiguation, and action reduces uncertainty through top-down contextual generation. Their closed-loop interaction aligns latent content variables across modalities, maximizing mutual information \( I(\Phi_s; \Phi_a) \) and enabling adaptive sensorimotor learning through iterative, embodied inference. 
\end{observation}

\textbf{Remark.} In summary, CCUP reconceptualizes the core PAC challenge, not as mere coordination, but as an information-theoretic alignment problem. By minimizing joint uncertainty and maximizing mutual information across asymmetrical flows, CCUP suggests a unifying principle for robust, generalizable sensorimotor learning.

\begin{tcolorbox}[title=Losing Balance: Highly Superior Autobiographical Memory (HSAM), colback=gray!5, colframe=blue, fonttitle=\bfseries]

\textbf{Definition:} Highly Superior Autobiographical Memory (HSAM) refers to the rare ability to recall personal past experiences with exceptional detail, accuracy, and temporal specificity, often dating back decades.

\textbf{Key Characteristics:}
\begin{itemize}
    \item \textbf{Exceptional episodic recall:} Individuals with HSAM can effortlessly retrieve detailed personal events for almost any date in their lives.
    \item \textbf{Involuntary retrieval:} Memories are often spontaneously triggered, not effortfully reconstructed.
    \item \textbf{Domain specificity:} The superior recall is limited to autobiographical events; performance on other memory tasks (e.g., working memory, abstract reasoning) is often normal.
    \item \textbf{Calendar anchoring:} HSAM individuals frequently anchor memories to calendar dates, suggesting strong time-context bindings.
\end{itemize}

\textbf{Neurobiological Correlates:}
\begin{itemize}
    \item \textbf{Structural differences:} Enlarged caudate nucleus and increased white matter density in regions associated with autobiographical memory.
    \item \textbf{Default mode network activity:} Some evidence suggests altered resting-state connectivity patterns, especially in regions involved in self-referential processing.
\end{itemize}

\textbf{Cognitive Trade-Offs:}
\begin{itemize}
    \item HSAM may involve reduced generalization due to over-retention of context-specific details.
    \item Some individuals report obsessive traits or heightened emotional salience attached to past events.
\end{itemize}


\end{tcolorbox}

\subsection{Memory Consolidation: Sleep–Wake Cycle}
\label{sec:4.3}

The Context–Content Uncertainty Principle (CCUP) naturally extends beyond immediate, online inference into offline cognitive processes such as memory consolidation \cite{stickgold2005sleep}. The biological sleep–wake cycle provides a critical substrate for slow, recursive alignment between content and context representations operating over hierarchical timescales \cite{fuller2006neurobiology}. While wakefulness prioritizes rapid, real-time cycle formation through active perception and action, sleep enables the brain to engage in extensive reprocessing of prior experiences by replaying episodic traces and abstracting across distributed contextual information \cite{wilson1994reactivation, diekelmann2010memory}.

From the CCUP perspective, memory consolidation during sleep can be conceptualized as an offline process of \emph{contextual disambiguation} and \emph{reconstruction}. Episodic memories, which are initially encoded within highly detailed, high-dimensional, and time-specific contextual frames, are reactivated and replayed without concurrent sensory input \cite{mcclelland1995there}. This replay is not a mere passive re-experience; rather, it actively leverages the reuse and aliasing of content representations across multiple contextual episodes, allowing the system to detect and extract structural regularities and invariant features embedded within noisy episodic data. Through this iterative inference process, the brain transforms high-entropy, context-rich episodic memories into more compressed, low-entropy semantic memories that are generalized and abstracted \cite{tse2007schemas}.
Critically, this memory consolidation mechanism minimizes the conditional entropies \( H(\Phi \mid \Psi) \) and \( H(\Psi \mid \Phi) \) at a slower, offline timescale, effectively aligning episodic content and contextual frames through cycle-consistent replay \cite{wilson1994reactivation}. This recursive alignment supports the gradual integration of episodic experiences into structured, semantic knowledge bases, contributing to lifelong learning and flexible cognition.

Moreover, this consolidation process forms part of a \emph{hierarchical CCUP architecture}: fast cycles govern online perception–action loops, enabling rapid inference and immediate behavioral adaptation, while slower cycles during sleep support the abstraction, integration, and accumulation of knowledge over extended periods \cite{buzsaki2006rhythms, hasson2015hierarchical}. These nested temporal dynamics facilitate a functional continuum from transient, task-specific encoding to stable, generalized representations. By orchestrating cycle formation across multiple temporal scales, CCUP reconceptualizes memory as an active, dynamic mapping from entropic experiential traces to structured latent representations, rather than a static storage system \cite{mcclelland1995there, hasson2015hierarchical}.

This perspective is supported by empirical evidence that sleep facilitates the reorganization of hippocampal–neocortical interactions \cite{buzsaki1996hippocampo}, promoting the transfer of detailed episodic information to more abstract cortical circuits \cite{stickgold2005sleep, diekelmann2010memory}. Through this interplay, the brain not only consolidates memories but also restructures internal models to optimize future inference and behavior, embodying CCUP’s principle of uncertainty reduction via cycle-consistent alignment. However, it should be noted that the study of Highly Superior Autobiographical Memory (HSAM) \cite{leport2012behavioral} suggests that the Caudate is often abnormally enlarged or hyperactive, contributing to a shifted balance from generalization to specification. While the macrostructure of sleep (e.g., REM/NREM cycles) is normal, the computational content of sleep-based memory processing may differ, favoring detailed reinstatement over generalization. Such observation inspires us to make the following observation.

\begin{observation}[Caudate-Induced Content-Context Overbinding in HSAM]
In individuals with photographic memory, the Caudate nucleus facilitates hyper-specific encoding by tightly binding content \( \Phi \) to unique, high-entropy contexts \( \Psi \), such that:
$H(\Phi \mid \Psi) \to 0, \quad H(\Psi \mid \Phi) \gg 0$.
This imbalance violates the CCUP-driven requirement for bidirectional alignment and generalization, leading to overfitted episodic representations and impaired semantic abstraction.
\end{observation}

\textbf{Remark.} Under the CCUP framework, HSAM may result from \emph{overbinding} of content \( \Phi \) to specific context \( \Psi \), potentially mediated by the caudate nucleus. This leads to an excessive minimization of conditional entropy \( H(\Phi \mid \Psi) \), producing context-saturated memory traces that resist abstraction. In typical cognition, generalization relies on \emph{contextual aliasing}—the reuse of the same \( \Phi \) across multiple \( \Psi \)'s—but in HSAM, this aliasing is disrupted, inflating \( H(\Psi \mid \Phi) \) and impairing abstraction. The bidirectional inference cycle between \( \Phi \) and \( \Psi \) becomes unbalanced: bottom-up content binding is overly rigid and detailed, while top-down contextual reconstruction is weak or absent. Consequently, the dynamic cycle that would normally resolve the IB through iterative refinement fails to operate effectively, preventing the system from reorganizing episodic memory into abstract, semantic structures. The enlarged caudate nucleus in HSAM \cite{leport2012behavioral} likely enhances associative binding, attentional selection, and temporal encoding, supporting the structured, compulsively rehearsed autobiographical memory seen in these individuals. It plays a complementary role to the hippocampus, focusing less on spatial-episodic binding and more on procedural and contextual reinforcement of personal history.

\subsection{Origin of Language as Cross-Brain Cycle Formation}

Within the CCUP framework, language can be understood as an evolutionary extension of the perception-action cycle, transforming internal sensorimotor inference into a socially shared inference mechanism. At its core, language enables the transport of latent content between agents by projecting internally simulated goals into a shared symbolic space. This projection allows two brains to form a coordinated inference cycle, resolving the IB in OT not only within individuals, but \emph{between} them.

\begin{tcolorbox}[colback=gray!5!white, colframe=blue!75!white, title=Case Study: Helen Keller and Delta-Seeded Hierarchical Inference]
\textbf{How does a child, blind and deaf from infancy, acquire language?} Helen Keller’s remarkable case can be understood as an extreme instance of learning under a severely constrained IB. Deprived of auditory and visual input, Keller relied almost exclusively on the tactile modality---a channel with high ambiguity and limited contextual richness. Yet through the guidance of her teacher, Anne Sullivan, she managed to form structured internal representations and develop symbolic communication. 
Within the CCUP framework, Keller's learning exemplifies \emph{delta-seeded hierarchical inference}. Each tactile spelling episode (e.g., ``W-A-T-E-R'' traced onto her palm) served as a delta-seeded symbolic attractor, a sharp, low-entropy latent representation. When this symbolic input was paired with a somatic experience (e.g., water running over her hand), the brain instantiated a co-occurring context that formed a cycle-consistent inference loop: tactile signal $\rightarrow$ latent concept $\rightarrow$ somatic context, and vice versa. This bootstrapped content binding and context disentangling, allowing Keller to infer specific symbolic meanings despite sensory aliasing.
Crucially, Keller’s cognitive architecture generalized these delta-seeded latents across novel tactile and environmental contexts, a diffusion process that expanded her internal semantic space. Over time, her growing symbolic system acted as a top-down prior, shaping selective inference at every level of her cognitive hierarchy. Keller’s case reveals how the brain can reconstruct abstract language through delta-anchored cycle formation and selective binding even in the absence of sight or sound.
\end{tcolorbox}

\paragraph{From Internal Inference to Social Projection.}
Within individual organisms, the PAC functions as a tightly coupled closed-loop system: internally generated goals and intentions act as \emph{delta seeds} that initiate motor actions, which in turn generate contextual sensory feedback. This feedback continuously updates the agent's latent representations, enabling iterative refinement of both content and context under the CCUP. Through this dynamic interplay, the system efficiently aligns low-entropy, goal-specific content with high-entropy, often ambiguous contextual information, thereby optimizing inference and behavior.
The emergence of language marks a profound evolutionary extension of this inferential architecture from the internal domain to the social realm \cite{bennett2023brief}. Rather than directly manipulating the physical environment, agents externalize their internal latent content by producing symbolic tokens—words, gestures, and vocalizations—that function as compressed, portable carriers of meaning. These symbols are crafted to trigger analogous inference cycles within other agents, effectively inducing corresponding latent content reconstruction and contextual disambiguation in listeners or observers.

From the CCUP perspective, language can be understood as a form of \emph{socially extended optimal transport} across minds, where symbolic communication enables the transmission and alignment of latent representations between individuals \cite{frith2007social}. This social projection transforms the closed internal inference cycle into an open, interactive system of recursive exchanges, allowing cooperative knowledge building, shared intentionality, and cultural evolution. Symbols thus serve not only as content seeds but also as mediators that shape the joint contextual space, supporting coordinated action and collective understanding \cite{robbins2005social}.
This framework highlights the dual role of language as both a tool for content binding within an individual’s cognitive system and a mechanism for context disentangling across social networks. By embedding individual inference cycles within larger communicative loops, CCUP provides a principled account of how language enables minds to extend their inferential capacities beyond the boundaries of the body and brain, fostering complex social cognition and shared meaning.

\paragraph{Language as Intersubjective Optimal Transport.}
Consider two agents engaged in communication: a speaker \( A \) with latent intention \( z_A \), and a listener \( B \) possessing contextual state \( x_B \) that may differ from the speaker's own context \( x_A \). The speaker emits a symbolic token \( u \), such as a word or gesture, which serves as a \emph{transport seed} facilitating information transfer across distinct cognitive systems controlled by the perception-action cycles of two agents.
Formally, this process can be represented as a chain of mappings:
$z_A \xrightarrow{\text{encoding}} u \xrightarrow{\text{decoding}} \hat{z}_B$,
where \( \hat{z}_B \) is the listener's inferred latent representation approximating the speaker's original intention \( z_A \). Crucially, the communication channel mediated by the symbol \( u \) implements a \emph{delta-seeded inference cycle} within the listener’s mind, enabling the reconstruction of the speaker’s intended content despite potentially differing contextual backgrounds.

The symbolic token \( u \) thus acts as an \emph{alignment operator}, designed to reduce uncertainty in the listener's inference by simultaneously lowering:
$H(z_A \mid u) \quad \text{and} \quad H(z_A \mid x_B)$,
thereby facilitating the resolution of the intersubjective \emph{information bottleneck} that arises due to asymmetries in context and perceptual experience.
Within the CCUP framework, language can therefore be interpreted as a specialized form of \emph{optimal transport} operating across minds. The symbol \( u \) provides a compact, low-entropy anchor (delta seed) that effectively transports high-dimensional, high-entropy intentions through a shared symbolic space. This social transport enables recursive inference cycles in the listener that dynamically disambiguate context and bind content, ensuring that meaning is preserved and adapted despite contextual divergence.
Furthermore, this intersubjective transport process supports the emergence of shared conceptual spaces and common ground, enabling cooperative behavior, cultural transmission, and collective intelligence. By externalizing internal latent states into symbolic forms, agents extend their inferential reach beyond individual cognition, harnessing language as a mechanism for joint epistemic alignment and social coordination.

\paragraph{Cycle Formation Across Brains.}
Language uniquely enables two agents to engage in a recursive process of mutual inference and refinement of internal states through symbolic exchange. Consider a speaker \( A \) with latent intention \( z_A \), who emits a symbol \( u \) to a listener \( B \). The listener decodes this symbol to form an inferred representation \( \hat{z}_B \), which then generates a response symbol \( u' \) sent back to the speaker. The speaker, in turn, decodes \( u' \) into \( \hat{z}_A \), closing the loop:
\[
z_A \xrightarrow{\text{encode}} u \xrightarrow{\text{decode}} \hat{z}_B \xrightarrow{\text{encode}} u' \xrightarrow{\text{decode}} \hat{z}_A.
\]

This recursive exchange constitutes a \emph{cross-brain cycle}, where both agents iteratively simulate, predict, and update each other's internal latent states through bidirectional symbolic communication \cite{friston2015active2}. Such cycles foster dynamic alignment of goals, beliefs, and inter-person contextual models by bootstrapping intra-person perception-action cycles, enabling \emph{shared intentionality}—the coordinated understanding and pursuit of common objectives.

From an information-theoretic viewpoint grounded in the CCUP, the cross-brain cycle functions to minimize joint uncertainty across the coupled latent spaces of both agents. By continuously updating predictions and incorporating feedback, the cycle reduces conditional entropies \( H(z_A \mid \hat{z}_B) \) and \( H(z_B \mid \hat{z}_A) \), enhancing mutual information and ensuring consistency between internal goals and the external communicative context.
Importantly, this intersubjective cycle enables \emph{collaborative inference}—the distributed construction of shared meaning that transcends individual cognitive limitations. It supports negotiation, error correction, and the emergence of cultural conventions, allowing language to function not simply as a static code but as a dynamic, adaptive process of joint cognition across minds \cite{clark1996language}.
Thus, cycle formation across brains via language exemplifies a biologically and socially grounded mechanism for extending the CCUP framework beyond individual cognition, highlighting the recursive, cyclic nature of communication and shared understanding in human social systems.

\paragraph{Bootstrapping Social Intelligence.}
Language fundamentally transforms social cognition by encoding complex, high-dimensional latent cognitive states into low-dimensional symbolic tokens. The invention of natural languages by nature \cite{bennett2023brief} enables several key capabilities:
1) \textbf{Efficient transmission of goals and abstract concepts}: Symbols act as compact carriers of intent and meaning, allowing agents to share complex plans, emotions, and ideas rapidly and reliably across noisy and variable social environments; 2) \textbf{Selective inference in the face of contextual ambiguity}: By providing precise delta-like seeds for inference, symbolic communication helps listeners disambiguate their own high-entropy social contexts, narrowing down possible interpretations to those relevant for shared understanding; 3) \textbf{Alignment of internal models across socially embedded agents}: Language scaffolds the recursive synchronization of beliefs, desires, and expectations, promoting convergence toward common ground and coordinated behavior despite differing individual experiences or perspectives; 4) \textbf{Recursive reasoning and counterfactual simulation across minds}: Through nested symbolic exchanges, agents engage in higher-order social cognition such as theory of mind \cite{baker2017rational}, perspective taking, and imagining alternative scenarios, enabling complex cooperation and cultural evolution.
Within the CCUP framework, language thus emerges as a biologically and cognitively grounded mechanism that extends the principles of delta-seeded diffusion, originally formulated for sensorimotor learning, into the social realm. This extension allows agents to collaboratively construct and refine shared cognitive spaces through recursive symbolic inference, effectively bootstrapping social intelligence from individual internal models. 

\begin{observation}[Recursive Theory of Mind via Symbolic Inference]
Let agents \( A \) and \( B \) participate in a cross-brain inference cycle mediated by symbolic communication under the CCUP framework. If agent \( A \) can simulate not only the latent state \( z_B \) of agent \( B \), but also agent \( B \)'s inferred model of \( A \)'s state \( \hat{z}_A \), then: 1) Agent \( A \)'s inference cycle includes a recursively embedded model: \( A \models B \models A \), enabling higher-order intentionality (\( A \models B \) means Agent A has a model of Agent B - i.e., A simulates B's mental states); 2) Symbolic utterances \( u \) and \( u' \) exchanged between the agents act as externalized delta seeds that scaffold this recursion, grounding counterfactual inference and anticipatory alignment; 3) This recursive embedding minimizes the joint information bottleneck \( IB(A \leftrightarrow B) \), allowing agents to achieve mutual modeling under uncertainty with sublinear communication cost.
\end{observation}

\textbf{Remark.} Language emerged as a cognitive mechanism for aligning recursive context-content inference across agents. Its hierarchical structure reflects internal cycles of Theory of Mind (ToM)-based simulation \cite{premack1978does}, encoded via syntax \cite{fitch2011evolution} to minimize communicative uncertainty. Thus, the generative power of language corresponds to the recursive depth of symbolic inference required for social reasoning. The CCUP theory suggests that language is the externalization of internal inference cycles used for recursive social reasoning. Syntax reflects the compression of memory chains and symbolic inferences into a communicable format, whose computational structure (Chomsky hierarchy \cite{chomsky1956three}) mirrors the depth of cognitive recursion required for ToM \cite{hauser2002faculty}.

\section{Conclusion: The Cyclical Nature of Inference}
\label{sec:5}

The CCUP reframes inference as a directional, hierarchical process shaped by fundamental asymmetries in information structure. Analogous to the arrow of time’s unidirectional increase in entropy, the arrow of inference arises from the dynamic interplay between high-entropy, ambiguous context and low-entropy, specific content. Inference proceeds through cyclic interactions that iteratively disambiguate context and reconstruct it from stable content representations, resolving an intrinsic information bottleneck via bidirectional alignment.
We have demonstrated that this cyclic architecture manifests across multiple cognitive domains and timescales. 
Importantly, these local inference cycles extend into structured \emph{memory chains}, where each cycle contributes to a temporally unfolding trajectory of latent states. This chain formation enables recursive goal simulation, counterfactual reasoning, and long-horizon policy refinement within a unified uncertainty-minimizing framework. During offline consolidation, sleep–wake cycles instantiate slower, hierarchical chains of inference that transform episodic experiences into abstract semantic knowledge through recurrent replay and dynamic context–content realignment across both temporal and spatial dimensions.


Extending beyond individual cognition, CCUP naturally accounts for social inference and language as forms of \emph{intersubjective optimal transport}. Language externalizes internal latent content as symbolic tokens that seed recursive inference cycles across agents, forming cross-brain cycles that support shared intentionality, collaborative inference, and cultural evolution. This socially extended cycle formation enables minds to dynamically simulate and align internal states, transforming language from a static code into an adaptive, interactive process of joint cognition.
Ultimately, CCUP provides a unifying framework that integrates perception, action, memory,  and social communication into a single principle: {\bf cognition is a hierarchy of cycle-consistent inference processes that minimize joint uncertainty by continually aligning content and context}. This arrow of inference, like the arrow of time, is not merely an emergent property but a structural necessity underlying intelligence and sociality.

\bibliographystyle{plain}
\bibliography{references}  

\begin{thebibliography}{100}

\bibitem{ajzen1986prediction}
Icek Ajzen and Thomas~J Madden.
\newblock Prediction of goal-directed behavior: Attitudes, intentions, and perceived behavioral control.
\newblock {\em Journal of experimental social psychology}, 22(5):453--474, 1986.

\bibitem{anderson1972more}
Philip~W Anderson.
\newblock More is different: Broken symmetry and the nature of the hierarchical structure of science.
\newblock {\em Science}, 177(4047):393--396, 1972.

\bibitem{baird2019cognitive}
Benjamin Baird, Sergio~A Mota-Rolim, and Martin Dresler.
\newblock The cognitive neuroscience of lucid dreaming.
\newblock {\em Neuroscience \& Biobehavioral Reviews}, 100:305--323, 2019.

\bibitem{baker2017rational}
Chris~L. Baker, Rebecca Saxe, and Joshua~B. Tenenbaum.
\newblock Rational quantitative attribution of beliefs, desires and percepts in human mentalizing.
\newblock {\em Nature Human Behaviour}, 1:1--10, 2017.

\bibitem{ballard1981generalizing}
Dana~H Ballard.
\newblock Generalizing the hough transform to detect arbitrary shapes.
\newblock {\em Pattern recognition}, 13(2):111--122, 1981.

\bibitem{bansal2023cold}
Anirudh Bansal, Tan Nguyen, Ziyan Wang, and Shiyu Chang.
\newblock Cold diffusion: Inverting arbitrary image corruptions without noise.
\newblock {\em arXiv preprint arXiv:2302.00275}, 2023.

\bibitem{bar2004visual}
Moshe Bar.
\newblock Visual objects in context.
\newblock {\em Nature Reviews Neuroscience}, 5(8):617--629, 2004.

\bibitem{barlow1972single}
Horace~B. Barlow.
\newblock Single units and sensation: a neuron doctrine for perceptual psychology?
\newblock {\em Perception}, 1(4):371--394, 1972.

\bibitem{bastos2012canonical}
Andre~M Bastos, W~Martin Usrey, Rick~A Adams, George~R Mangun, Pascal Fries, and Karl~J Friston.
\newblock Canonical microcircuits for predictive coding.
\newblock {\em Neuron}, 76(4):695--711, 2012.

\bibitem{bellman1966dynamic}
Richard Bellman.
\newblock Dynamic programming.
\newblock {\em science}, 153(3731):34--37, 1966.

\bibitem{bengio2013representation}
Yoshua Bengio, Aaron Courville, and Pascal Vincent.
\newblock Representation learning: A review and new perspectives.
\newblock In {\em IEEE transactions on pattern analysis and machine intelligence}, volume~35, pages 1798--1828. IEEE, 2013.

\bibitem{bennett2023brief}
Max Bennett.
\newblock {\em A brief history of intelligence: evolution, AI, and the five breakthroughs that made our brains}.
\newblock HarperCollins, 2023.

\bibitem{beny2013quantum}
C{\'e}dric B{\'e}ny.
\newblock Deep learning and the renormalization group.
\newblock {\em arXiv preprint arXiv:1301.3124}, 2013.

\bibitem{blackwell1947conditional}
David Blackwell.
\newblock Conditional expectation and unbiased sequential estimation.
\newblock {\em The Annals of Mathematical Statistics}, pages 105--110, 1947.

\bibitem{blundell2015weight}
Charles Blundell, Julien Cornebise, Koray Kavukcuoglu, and Daan Wierstra.
\newblock Weight uncertainty in neural networks.
\newblock In {\em ICML}, 2015.

\bibitem{botvinick2020deep}
Matthew Botvinick, Sam Ritter, Jane~X Wang, Zeb Kurth-Nelson, Charles Blundell, and Demis Hassabis.
\newblock Deep reinforcement learning and its neuroscientific implications.
\newblock {\em Neuron}, 107(4):603--616, 2020.

\bibitem{bousquet2002stability}
Olivier Bousquet and André Elisseeff.
\newblock Stability and generalization.
\newblock {\em Journal of Machine Learning Research}, 2:499--526, 2002.

\bibitem{buzsaki1996hippocampo}
Gy{\"o}rgy Buzs{\'a}ki.
\newblock The hippocampo-neocortical dialogue.
\newblock {\em Cerebral cortex}, 6(2):81--92, 1996.

\bibitem{buzsaki2006rhythms}
Gy{\"o}rgy Buzs{\'a}ki.
\newblock {\em Rhythms of the Brain}.
\newblock Oxford university press, 2006.

\bibitem{chomsky1956three}
Noam Chomsky.
\newblock Three models for the description of language.
\newblock {\em IRE Transactions on information theory}, 2(3):113--124, 1956.

\bibitem{clark2013whatever}
Andy Clark.
\newblock Whatever next? predictive brains, situated agents, and the future of cognitive science.
\newblock {\em Behavioral and Brain Sciences}, 36(3):181--204, 2013.

\bibitem{clark1996language}
Herbert~H. Clark.
\newblock {\em Using Language}.
\newblock Cambridge University Press, Cambridge, UK, 1996.

\bibitem{colgin2008understanding}
Laura~Lee Colgin, Edvard~I Moser, and May-Britt Moser.
\newblock Understanding memory through hippocampal remapping.
\newblock {\em Trends in neurosciences}, 31(9):469--477, 2008.

\bibitem{cover1999elements}
Thomas~M Cover.
\newblock {\em Elements of information theory}.
\newblock John Wiley \& Sons, 1999.

\bibitem{dayan1995helmholtz}
Peter Dayan, Geoffrey~E. Hinton, Radford~M. Neal, and Richard~S. Zemel.
\newblock The helmholtz machine.
\newblock {\em Neural Computation}, 7(5):889--904, 1995.

\bibitem{dicarlo2007untangling}
James~J DiCarlo and David~D Cox.
\newblock Untangling invariant object recognition.
\newblock {\em Trends in cognitive sciences}, 11(8):333--341, 2007.

\bibitem{dicarlo2012does}
James~J DiCarlo, Davide Zoccolan, and Nicole~C Rust.
\newblock How does the brain solve visual object recognition?
\newblock {\em Neuron}, 73(3):415--434, 2012.

\bibitem{diekelmann2010memory}
Susanne Diekelmann and Jan Born.
\newblock The memory function of sleep.
\newblock {\em Nature Reviews Neuroscience}, 11(2):114--126, 2010.

\bibitem{esmaeili2019structured}
Behzad Esmaeili, Hao Wu, Shikhar Jain, Brooks Paige, Ioannis Mitliagkas, and Richard Zemel.
\newblock Structured disentangled representations.
\newblock In {\em Proceedings of the 33rd International Conference on Neural Information Processing Systems}, 2019.

\bibitem{fitch2011evolution}
W~Tecumseh Fitch.
\newblock The evolution of syntax: an exaptationist perspective.
\newblock {\em Frontiers in evolutionary neuroscience}, 3:9, 2011.

\bibitem{friston2005theory}
Karl Friston.
\newblock A theory of cortical responses.
\newblock {\em Philosophical Transactions of the Royal Society B: Biological Sciences}, 360(1456):815--836, 2005.

\bibitem{friston2006free}
Karl Friston.
\newblock A free energy principle for the brain.
\newblock {\em Journal of Physiology-Paris}, 100(1–3):70--87, 2006.

\bibitem{friston2008hierarchical}
Karl Friston.
\newblock Hierarchical models in the brain.
\newblock {\em PLoS Computational Biology}, 4(11):e1000211, 2008.

\bibitem{friston2010free}
Karl Friston.
\newblock The free-energy principle: a unified brain theory?
\newblock {\em Nature Reviews Neuroscience}, 11(2):127--138, 2010.

\bibitem{friston2017graphical}
Karl Friston, Thomas Parr, and Bert de~Vries.
\newblock Graphical brain: Belief propagation and active inference.
\newblock {\em Network Neuroscience}, 1(4):381--414, 2017.

\bibitem{friston2017active}
Karl Friston, Francesco Rigoli, David Ognibene, Christoph Mathys, Thomas FitzGerald, Giovanni Pezzulo, and ...
\newblock Active inference: a process theory.
\newblock {\em Neural Computation}, 29(1):1--49, 2017.

\bibitem{friston2015active1}
Karl Friston, Richard Rosch, Thomas Parr, Christopher Price, and Howard Bowman.
\newblock Active inference and epistemic value.
\newblock {\em Cognitive Neuroscience}, 6(4):187--214, 2015.

\bibitem{friston2010action}
Karl~J Friston, Jean Daunizeau, James Kilner, and Stefan~J Kiebel.
\newblock Action and behavior: a free-energy formulation.
\newblock {\em Biological cybernetics}, 102:227--260, 2010.

\bibitem{friston2015active2}
Karl~J Friston and Christopher~D Frith.
\newblock Active inference, communication and hermeneutics.
\newblock {\em cortex}, 68:129--143, 2015.

\bibitem{frith2007social}
Chris~D. Frith and Uta Frith.
\newblock The social brain?
\newblock {\em Philosophical Transactions of the Royal Society B: Biological Sciences}, 362(1480):671--678, 2007.

\bibitem{fuller2006neurobiology}
Patrick~M Fuller, Joshua~J Gooley, and Clifford~B Saper.
\newblock Neurobiology of the sleep-wake cycle: sleep architecture, circadian regulation, and regulatory feedback.
\newblock {\em Journal of biological rhythms}, 21(6):482--493, 2006.

\bibitem{fuster2004upper}
Joaquin~M. Fuster.
\newblock Upper processing stages of the perception–action cycle.
\newblock {\em Trends in Cognitive Sciences}, 8(4):143--145, 2004.

\bibitem{geman1992neural}
Stuart Geman, Elie Bienenstock, and Ren{\'e} Doursat.
\newblock Neural networks and the bias/variance dilemma.
\newblock {\em Neural computation}, 4(1):1--58, 1992.

\bibitem{geman1984stochastic}
Stuart Geman and Donald Geman.
\newblock Stochastic relaxation, gibbs distributions, and the bayesian restoration of images.
\newblock {\em IEEE Transactions on pattern analysis and machine intelligence}, (6):721--741, 1984.

\bibitem{george2021clone}
Dileep George, Rajeev~V Rikhye, Nishad Gothoskar, J~Swaroop Guntupalli, Antoine Dedieu, and Miguel L{\'a}zaro-Gredilla.
\newblock Clone-structured graph representations enable flexible learning and vicarious evaluation of cognitive maps.
\newblock {\em Nature communications}, 12(1):2392, 2021.

\bibitem{gershman2019what}
Samuel~J. Gershman.
\newblock What does the free energy principle tell us about the brain?
\newblock {\em Trends in Cognitive Sciences}, 23(10):785--797, 2019.

\bibitem{gibson1979ecological}
James~J. Gibson.
\newblock The ecological approach to visual perception: Classic edition.
\newblock 1979.

\bibitem{goodale1992separate}
Melvyn~A. Goodale and A.~David Milner.
\newblock Separate visual pathways for perception and action.
\newblock {\em Trends in Neurosciences}, 15(1):20--25, 1992.

\bibitem{goyal2022inductive}
Anirudh Goyal and Yoshua Bengio.
\newblock Inductive biases for deep learning of higher-level cognition.
\newblock {\em Proceedings of the Royal Society A}, 478(2266):20210068, 2022.

\bibitem{hasson2015hierarchical}
Uri Hasson, Janice Chen, and Christopher~J Honey.
\newblock Hierarchical process memory: memory as an integral component of information processing.
\newblock {\em Trends in cognitive sciences}, 19(6):304--313, 2015.

\bibitem{hauser2002faculty}
Marc~D Hauser, Noam Chomsky, and W.~Tecumseh Fitch.
\newblock The faculty of language: What is it, who has it, and how did it evolve?
\newblock {\em Science}, 298(5598):1569--1579, 2002.

\bibitem{higgins2017beta}
Irina Higgins, Loic Matthey, Arka Pal, Christopher Burgess, Xavier Glorot, Matthew Botvinick, Shakir Mohamed, and Alexander Lerchner.
\newblock beta-vae: Learning basic visual concepts with a constrained variational framework.
\newblock {\em International Conference on Learning Representations (ICLR)}, 2017.

\bibitem{hinton1995wake}
Geoffrey~E Hinton, Peter Dayan, Brendan~J Frey, and Radford~M Neal.
\newblock The" wake-sleep" algorithm for unsupervised neural networks.
\newblock {\em Science}, 268(5214):1158--1161, 1995.

\bibitem{ho2020denoising}
Jonathan Ho, Ajay Jain, and Pieter Abbeel.
\newblock Denoising diffusion probabilistic models.
\newblock {\em Advances in Neural Information Processing Systems (NeurIPS)}, 33:6840--6851, 2020.

\bibitem{ho2022classifierfree}
Jonathan Ho and Tim Salimans.
\newblock Classifier-free diffusion guidance.
\newblock In {\em arXiv preprint arXiv:2207.12598}, 2022.

\bibitem{hullermeier2021aleatoric}
Eyke H{\"u}llermeier and Willem Waegeman.
\newblock Aleatoric and epistemic uncertainty in machine learning: An introduction to concepts and methods.
\newblock {\em Machine learning}, 110(3):457--506, 2021.

\bibitem{ihler2005loopy}
Alexander~T. Ihler, John~W. Fisher~III, and Alan~S. Willsky.
\newblock Loopy belief propagation: Convergence and effects of message errors.
\newblock {\em Journal of Machine Learning Research}, 6:905--936, 2005.

\bibitem{izhikevich2006polychronization}
Eugene~M Izhikevich.
\newblock Polychronization: computation with spikes.
\newblock {\em Neural computation}, 18(2):245--282, 2006.

\bibitem{jeannerod1994representing}
Marc Jeannerod.
\newblock The representing brain: Neural correlates of motor intention and imagery.
\newblock {\em Behavioral and Brain sciences}, 17(2):187--202, 1994.

\bibitem{keller2018predictive}
Georg~B Keller and Thomas~D Mrsic-Flogel.
\newblock Predictive processing: a canonical cortical computation.
\newblock {\em Neuron}, 100(2):424--435, 2018.

\bibitem{kingma2014auto}
Diederik~P. Kingma and Max Welling.
\newblock Auto-encoding variational bayes.
\newblock {\em arXiv preprint arXiv:1312.6114}, 2014.

\bibitem{kingma2019introduction}
Diederik~P Kingma and Max Welling.
\newblock An introduction to variational autoencoders.
\newblock {\em Foundations and Trends in Machine Learning}, 12(4):307--392, 2019.

\bibitem{koller2009probabilistic}
Daphne Koller and Nir Friedman.
\newblock {\em Probabilistic Graphical Models: Principles and Techniques}.
\newblock MIT press, 2009.

\bibitem{leport2012behavioral}
Aurora~KR LePort, Aaron~T Mattfeld, Heather Dickinson-Anson, James~H Fallon, Craig~EL Stark, Frithjof Kruggel, Larry Cahill, and James~L McGaugh.
\newblock Behavioral and neuroanatomical investigation of highly superior autobiographical memory (hsam).
\newblock {\em Neurobiology of learning and memory}, 98(1):78--92, 2012.

\bibitem{li2025CCUP}
Xin Li.
\newblock On content-context uncertainty principle.
\newblock {\em Neural Information Processing Symposium}, 2025.
\newblock under review.

\bibitem{linsker1990perceptual}
Ralph Linsker.
\newblock Perceptual neural organization: Some approaches based on network models and information theory.
\newblock {\em Annual Review of Neuroscience}, 13(1):257--281, 1990.

\bibitem{locatello2019challenging}
Francesco Locatello, Stefan Bauer, Mario Lucic, Gunnar Raetsch, Sylvain Gelly, Bernhard Schoelkopf, and Olivier Bachem.
\newblock Challenging common assumptions in the unsupervised learning of disentangled representations.
\newblock In {\em ICML}, 2019.

\bibitem{malkov2018efficient}
Yu~A. Malkov and D.~A. Yashunin.
\newblock Efficient and robust approximate nearest neighbor search using hierarchical navigable small world graphs.
\newblock {\em IEEE Transactions on Pattern Analysis and Machine Intelligence}, 42(4):824--836, 2020.

\bibitem{mante2013context}
Valerio Mante, David Sussillo, Krishna~V. Shenoy, and William~T. Newsome.
\newblock Context-dependent computation by recurrent dynamics in prefrontal cortex.
\newblock {\em Nature}, 503(7474):78--84, 2013.

\bibitem{mcclelland1995there}
James~L. McClelland, Bruce~L. McNaughton, and Randall~C. O'Reilly.
\newblock Why there are complementary learning systems in the hippocampus and neocortex: Insights from the successes and failures of connectionist models of learning and memory.
\newblock {\em Psychological Review}, 102(3):419--457, 1995.

\bibitem{mehta2014exact}
Pankaj Mehta and David~J Schwab.
\newblock An exact mapping between the variational renormalization group and deep learning.
\newblock {\em arXiv preprint arXiv:1410.3831}, 2014.

\bibitem{millidge2021predictive}
Beren Millidge, Alexander Tschantz, Anil~K Seth, and Christopher~L Buckley.
\newblock Predictive coding: a theoretical and experimental review.
\newblock {\em Frontiers in Computational Neuroscience}, 15:17, 2021.

\bibitem{minsky1961steps}
Marvin Minsky.
\newblock Steps toward artificial intelligence.
\newblock {\em Proceedings of the IRE}, 49(1):8--30, 1961.

\bibitem{moser2008place}
Edvard~I Moser, Emilio Kropff, and May-Britt Moser.
\newblock Place cells, grid cells, and the brain's spatial representation system.
\newblock {\em Annu. Rev. Neurosci.}, 31(1):69--89, 2008.

\bibitem{olshausen2005toward}
Bruno~A. Olshausen, Li~Zhaoping, and James~C. Roddey.
\newblock Toward a unified theory of visual area v4.
\newblock {\em Visual Neuroscience}, 22(5):343--353, 2005.

\bibitem{parr2017uncertainty}
Thomas Parr and Karl~J. Friston.
\newblock Uncertainty, epistemics and active inference.
\newblock {\em Journal of the Royal Society Interface}, 14(136):20170376, 2017.

\bibitem{parr2022active}
Thomas Parr, Giovanni Pezzulo, and Karl Friston.
\newblock Active inference: the free energy principle in mind, brain, and behavior.
\newblock {\em MIT Press}, 2022.

\bibitem{parr2019generalised}
Thomas Parr, Giovanni Pezzulo, and Karl~J. Friston.
\newblock Generalised free energy and active inference: can the future cause the past?
\newblock {\em bioRxiv}, 2019.

\bibitem{pearl2009causality}
Judea Pearl.
\newblock Causality: models, reasoning and inference.
\newblock {\em Cambridge University Press}, 2009.

\bibitem{pearl2018book}
Judea Pearl and Dana Mackenzie.
\newblock {\em The book of why: the new science of cause and effect}.
\newblock Basic books, 2018.

\bibitem{pezzulo2014internally}
Giovanni Pezzulo, Matthijs~AA Van~der Meer, Carien~S Lansink, and Cyriel~MA Pennartz.
\newblock Internally generated sequences in learning and executing goal-directed behavior.
\newblock {\em Trends in cognitive sciences}, 18(12):647--657, 2014.

\bibitem{pickering2004toward}
Martin~J. Pickering and Simon Garrod.
\newblock Toward a mechanistic psychology of dialogue.
\newblock {\em Behavioral and Brain Sciences}, 27(2):169--190, 2004.

\bibitem{poggio2004generalization}
Tomaso Poggio and Emilio Bizzi.
\newblock Generalization in vision and motor control.
\newblock {\em Nature}, 431(7010):768--774, 2004.

\bibitem{premack1978does}
David Premack and Guy Woodruff.
\newblock Does the chimpanzee have a theory of mind?
\newblock {\em Behavioral and Brain Sciences}, 1(4):515--526, 1978.

\bibitem{quiroga2005invariant}
Rodrigo~Quian Quiroga, Lila Reddy, Gabriel Kreiman, Christof Koch, and Itzhak Fried.
\newblock Invariant visual representation by single neurons in the human brain.
\newblock {\em Nature}, 435(7045):1102--1107, 2005.

\bibitem{rao1992information}
C~Radhakrishna Rao.
\newblock Information and the accuracy attainable in the estimation of statistical parameters.
\newblock In {\em Breakthroughs in Statistics: Foundations and basic theory}, pages 235--247. Springer, 1992.

\bibitem{rao1999predictive}
Rajesh~PN Rao and Dana~H Ballard.
\newblock Predictive coding in the visual cortex: a functional interpretation of some extra-classical receptive-field effects.
\newblock {\em Nature Neuroscience}, 2(1):79--87, 1999.

\bibitem{robbins2005social}
Jordan~M Robbins and Joachim~I Krueger.
\newblock Social projection to ingroups and outgroups: A review and meta-analysis.
\newblock {\em Personality and social psychology review}, 9(1):32--47, 2005.

\bibitem{robertson1929uncertainty}
Howard~Percy Robertson.
\newblock The uncertainty principle.
\newblock {\em Physical Review}, 34(1):163, 1929.

\bibitem{rombach2022high}
Robin Rombach, Andreas Blattmann, Dominik Lorenz, Patrick Esser, and Björn Ommer.
\newblock High-resolution image synthesis with latent diffusion models.
\newblock {\em Proceedings of the IEEE/CVF Conference on Computer Vision and Pattern Recognition (CVPR)}, pages 10684--10695, 2022.

\bibitem{schacter2007cognitive}
Daniel~L Schacter and Donna~Rose Addis.
\newblock The cognitive neuroscience of constructive memory: remembering the past and imagining the future.
\newblock {\em Philosophical Transactions of the Royal Society B: Biological Sciences}, 362(1481):773--786, 2007.

\bibitem{shannon1948mathematical}
Claude~E Shannon.
\newblock A mathematical theory of communication.
\newblock {\em The Bell system technical journal}, 27(3):379--423, 1948.

\bibitem{shapiro2019embodied}
Lawrence Shapiro.
\newblock {\em Embodied cognition}.
\newblock Routledge, 2019.

\bibitem{sohn2021network}
Hansem Sohn, Nicolas Meirhaeghe, Rishi Rajalingham, and Mehrdad Jazayeri.
\newblock A network perspective on sensorimotor learning.
\newblock {\em Trends in Neurosciences}, 44(3):170--181, 2021.

\bibitem{song2021score}
Yang Song, Jascha Sohl-Dickstein, Diederik~P. Kingma, Abhishek Kumar, Stefano Ermon, and Ben Poole.
\newblock Score-based generative modeling through stochastic differential equations.
\newblock In {\em ICLR}, 2021.

\bibitem{stickgold2005sleep}
Robert Stickgold.
\newblock Sleep-dependent memory consolidation.
\newblock {\em Nature}, 437(7063):1272--1278, 2005.

\bibitem{sutton1998reinforcement}
Richard~S Sutton and Andrew~G Barto.
\newblock {\em Reinforcement Learning: An Introduction}.
\newblock MIT Press, 1998.

\bibitem{tishby2000information}
Naftali Tishby, Fernando Pereira, and William Bialek.
\newblock {\em The information bottleneck method}.
\newblock 2000.

\bibitem{tishby2011information}
Naftali Tishby and Daniel Polani.
\newblock Information bottleneck and statistical mechanics.
\newblock In {\em Proceedings of the 27th Conference on Uncertainty in Artificial Intelligence (UAI)}, 2011.

\bibitem{tishby2015deep}
Naftali Tishby and Noga Zaslavsky.
\newblock Deep learning and the information bottleneck principle.
\newblock In {\em IEEE Information Theory Workshop (ITW)}, pages 1--5. IEEE, 2015.

\bibitem{treisman1980feature}
Anne Treisman and Garry Gelade.
\newblock A feature-integration theory of attention.
\newblock {\em Cognitive Psychology}, 12(1):97--136, 1980.

\bibitem{tse2007schemas}
Dorothy Tse, Rosamund~F Langston, Masaki Kakeyama, Ingrid Bethus, Patrick~A Spooner, Emma~R Wood, Menno~P Witter, and Richard~GM Morris.
\newblock Schemas and memory consolidation.
\newblock {\em Science}, 316(5821):76--82, 2007.

\bibitem{tulving2002episodic}
Endel Tulving.
\newblock Episodic memory: From mind to brain.
\newblock {\em Annual review of psychology}, 53(1):1--25, 2002.

\bibitem{ungerleider1994and}
Leslie~G Ungerleider and James~V Haxby.
\newblock ‘what’and ‘where’in the human brain.
\newblock {\em Current opinion in neurobiology}, 4(2):157--165, 1994.

\bibitem{ungerleider2000mechanisms}
Leslie~G. Ungerleider and Sabine Kastner.
\newblock Mechanisms of visual attention in the human cortex.
\newblock {\em Annual Review of Neuroscience}, 23:315--341, 2000.

\bibitem{villani2009optimal}
Cédric Villani.
\newblock {\em Optimal Transport: Old and New}.
\newblock Springer, 2009.

\bibitem{wainwright2008graphical}
Martin~J Wainwright and Michael~I Jordan.
\newblock {\em Graphical Models, Exponential Families, and Variational Inference}.
\newblock Now Publishers Inc., 2008.

\bibitem{wang2018prefrontal}
Jane~X Wang, Zeb Kurth-Nelson, Dharshan Kumaran, Dhruva Tirumala, Hubert Soyer, Joel~Z Leibo, Demis Hassabis, and Matthew Botvinick.
\newblock Prefrontal cortex as a meta-reinforcement learning system.
\newblock {\em Nature Neuroscience}, 21(6):860--868, 2018.

\bibitem{weiss2001optimality}
Yair Weiss and William~T Freeman.
\newblock On the optimality of solutions of the max-product belief-propagation algorithm in arbitrary graphs.
\newblock {\em IEEE Transactions on information theory}, 47(2):736--744, 2001.

\bibitem{wilson1974renormalization}
Kenneth~G Wilson.
\newblock The renormalization group: Critical phenomena and the kondo problem.
\newblock {\em Reviews of Modern Physics}, 47(4):773--840, 1975.

\bibitem{wilson1994reactivation}
Matthew~A. Wilson and Bruce~L. McNaughton.
\newblock Reactivation of hippocampal ensemble memories during sleep.
\newblock {\em Science}, 265(5172):676--679, 1994.

\bibitem{wolpert2011principles}
Daniel~M Wolpert, J{\"o}rn Diedrichsen, and J~Randall Flanagan.
\newblock Principles of sensorimotor learning.
\newblock {\em Nature reviews neuroscience}, 12(12):739--751, 2011.

\bibitem{wolpert1995internal}
Daniel~M. Wolpert, Zoubin Ghahramani, and Michael~I. Jordan.
\newblock An internal model for sensorimotor integration.
\newblock {\em Science}, 269(5232):1880--1882, 1995.

\bibitem{zeh2007physical}
H.~Dieter Zeh.
\newblock {\em The Physical Basis of the Direction of Time}.
\newblock Springer, 5th edition, 2007.

\bibitem{zhang2017understanding}
Chiyuan Zhang, Samy Bengio, Moritz Hardt, Benjamin Recht, and Oriol Vinyals.
\newblock Understanding deep learning requires rethinking generalization.
\newblock {\em arXiv preprint arXiv:1611.03530}, 2017.

\end{thebibliography}

\newpage
\appendix

\section{Proof of Theorem 1}
\label{appendix:A}

\begin{proof}[Proof]
We start from the definition of mutual information:
\[
I(\Psi; \Phi) = H(\Phi) - H(\Phi \mid \Psi) = H(\Psi) - H(\Psi \mid \Phi).
\]

Rearranging both expressions:
\[
H(\Phi \mid \Psi) = H(\Phi) - I(\Psi; \Phi), \quad H(\Psi \mid \Phi) = H(\Psi) - I(\Psi; \Phi).
\]

Summing these two expressions:
\[
H(\Phi \mid \Psi) + H(\Psi \mid \Phi) = H(\Phi) + H(\Psi) - 2 I(\Psi; \Phi).
\]

Now, using the upper bound on mutual information:
\[
I(\Psi; \Phi) \leq \min\{H(\Phi), H(\Psi)\},
\]
we obtain the lower bound:
\[
H(\Phi \mid \Psi) + H(\Psi \mid \Phi) \geq H(\Phi) + H(\Psi) - 2 \min\{H(\Phi), H(\Psi)\}.
\]

This simplifies to:
\[
H(\Phi \mid \Psi) + H(\Psi \mid \Phi) \geq |H(\Phi) - H(\Psi)|.
\]

This bound captures a fundamental implication of CCUP: due to the inherent asymmetry in entropy between context and content, directional uncertainty cannot be jointly minimized beyond the entropy gap. The inequality is tight when mutual information saturates at \( \min\{H(\Phi), H(\Psi)\} \), corresponding to maximal representational alignment.
\end{proof}

\section{Proof of Theorem 2}
\label{appendix:B}

\begin{proof}
Let $X \sim \mu$, and define $Y := T(X) \sim \nu$. Since $T_{\#}\mu = \nu$, this defines a valid transport plan. Because $T$ is invertible almost everywhere under $\mu$ with inverse $T^{-1}$, the joint distribution $(X, Y)$ induces a deterministic mapping:
\[
X = T^{-1}(Y), \quad \mu\text{-almost surely}.
\]
This implies that $X$ is a deterministic function of $Y$, so the conditional entropy satisfies:
\[
H(X \mid Y) = 0.
\]
By the definition of mutual information:
\[
I(X; Y) = H(X) - H(X \mid Y) = H(X).
\]
Hence,
\[
\text{IB Loss} = H(X) - I(X; Y) = H(X) - H(X) = 0.
\]
Thus, the information bottleneck is resolved.
\end{proof}

\section{Proof of Theorem 3}
\label{appendix:C}

\begin{proof}[Sketch of Proof]
We sketch the argument in three main steps:

\textbf{1. Objective: CCUP-driven Inference Minimizes Joint Uncertainty.} \\
Under the Context-Content Uncertainty Principle (CCUP), each memory state $Z_t$ participates in a local inference cycle with its corresponding context $\Psi_t$ and content $\Phi_t$. The chain-level objective seeks to minimize uncertainty in both directions:
\[
\min_{q(Z_{1:T})} \sum_{t=1}^{T} \left[ H(Z_t \mid \Psi_t) + H(\Psi_t \mid Z_t) \right].
\]
Since $\Psi_t$ and $\Phi_t$ are latent projections of $Z_t$, this objective reduces to minimizing the conditional uncertainty within and across adjacent latent states:
\[
\min_{q(Z_{1:T})} \sum_{t=1}^{T} H(Z_t \mid Z_{t-1}) + H(Z_{t-1} \mid Z_t).
\]

\textbf{2. Approximate Inference via Structured LBP.} \\
The full posterior $q(Z_{1:T})$ is generally intractable due to loops formed by recurrent reconstruction and lateral associations. To address this, we adopt Loopy Belief Propagation (LBP) to iteratively approximate marginals $q_t(Z_t)$ and $q_{tt'}(Z_t, Z_{t'})$. LBP can be viewed as minimizing the Bethe Free Energy:
\[
\mathcal{F}_{\text{Bethe}} = \sum_t H(q_t) - \sum_{(t,t')} I(q_{tt'}),
\]
subject to local consistency constraints. These beliefs are updated by message passing:
\[
m_{s \to t}(Z_t) \propto \sum_{Z_s} \psi_{st}(Z_s, Z_t) \prod_{u \in \mathcal{N}(s) \setminus t} m_{u \to s}(Z_s),
\]
where $\psi_{st}$ are pairwise potentials derived from local context–content cycles.

\textbf{3. Convergence to Consistent Representations.} \\
Under mild conditions (weak loops, suitable damping), LBP converges to fixed points of the Bethe Free Energy, yielding self-consistent marginal beliefs:
\[
q(Z_t) \approx \text{argmin}_{q_t} \left[ H(q_t) - \sum_{t' \in \mathcal{N}(t)} I(q_{tt'}) \right].
\]
These beliefs reduce joint uncertainty across the memory chain, aligning with the CCUP principle that inference proceeds through dynamic consistency between context and content. Thus, memory retrieval, goal simulation, and consolidation emerge as inference over this latent chain, approximated by LBP.

\end{proof}

\section{Proof of Theorem 4}
\label{appendix:D}

\begin{proof}[Sketch of Proof]
We outline the argument in three main steps, emphasizing the equivalence between chain-based inference and renormalization group (RG) flows:

\textbf{1. Local Inference as Renormalization Steps.}\\
Consider a memory chain composed of latent states $\{(\Psi_t,\Phi_t)\}_{t=1}^{T}$, each pair minimizing local conditional entropy via structured Loopy Belief Propagation (LBP). Each inference cycle can be expressed as a local transformation:
$$
q(Z_{t-1}) \mapsto q(Z_t),
$$
where $q(Z_t)$ denotes the marginal belief state at step $t$. This transformation implicitly coarse-grains local details (irrelevant uncertainty) while preserving contextually relevant information. Thus, each inference step behaves analogously to a single RG transformation, reducing complexity while maintaining predictive structure.

\textbf{2. Chain Formation as Iterated RG Flow.}\\
Extending the local step to a chain of multiple inference cycles, we have a sequential composition of inference operations:
$$
q(Z_1) \mapsto q(Z_2) \mapsto \dots \mapsto q(Z_T).
$$
Each step represents a scale-dependent transformation of latent variables. Under CCUP, the chain systematically compresses uncertainty and refines representations. This process is formally analogous to iterated RG flows, wherein microscopic states (high-resolution initial beliefs) progressively evolve toward macroscopic descriptions (abstracted latent beliefs) across scales or steps.

\textbf{3. Convergence to a Fixed-Point RG Structure.}\\
As the chain stabilizes, the latent representations converge to a fixed-point distribution characterized by minimized irrelevant entropy and maximal preserved information:
$$
I(Z_t; Z_{t+1}) \approx \max.
$$
This fixed-point condition aligns with the critical points of RG flows, marking stable, scale-invariant representations. Consequently, the chain-based inference trajectory under CCUP effectively implements an RG flow over latent cognitive states, achieving both temporal consistency and hierarchical abstraction.

Thus, the cognitive inference implemented by structured LBP along a memory chain is formally equivalent to a renormalization group process, dynamically aligning latent representations at multiple spatial and temporal scales.

\end{proof}





\end{document}